\documentclass[showpacs,preprintnumbers,nofootinbib]{revtex4} 

\usepackage{graphicx}
\usepackage{dcolumn}
\usepackage{bm}
\usepackage{amssymb}
\usepackage{amsmath}
\usepackage{epsfig}    
\usepackage{color}    

\def\beq{\begin{equation}}
\def\eeq{\end{equation}}
\def\bea{\begin{array}}
\def\eea{\end{array}}
\def\be{\begin{equation}}
\def\ee{\end{equation}}
\def\ba{\begin{eqnarray}}
\def\ea{\end{eqnarray}}

\def\to{\rightarrow}

\def\[{\left[}
\def\]{\right]}
\def\({\left(}
\def\){\right)}

\def\sm0{{\widetilde{m}_0}}

\def\U1em{{U(1)_{\rm em}}}
\def\to{\rightarrow}

\def\sq2{\sqrt{2}}

\def\ee{e^+e^-}

\def\End{\end{document}}

\def\fsl#1{\setbox0=\hbox{$#1$}                 
   \dimen0=\wd0                                 
   \setbox1=\hbox{/} \dimen1=\wd1               
   \ifdim\dimen0>\dimen1                        
      \rlap{\hbox to \dimen0{\hfil/\hfil}}      
      #1                                        
   \else                                        
      \rlap{\hbox to \dimen1{\hfil$#1$\hfil}}   
      /                                         
   \fi}

\begin{document} 

\title{Radiative corrections to electroweak parameters in the Higgs triplet model and 
implication with the recent Higgs boson searches}%
\preprint{UT-HET 064}
\author{%
{\sc Shinya Kanemura\,,   
     Kei Yagyu\,}}
\affiliation{
Department of Physics, University of Toyama, 3190 Gofuku, Toyama 930-8555, Japan\\}
\begin{abstract}
We study radiative corrections to the 
electroweak parameters in the Higgs model with the $Y=1$ triplet field, 
which is introduced in the scenario of generating neutrino masses based on the so-called type II seesaw mechanism. 
In this model, the rho parameter deviates from unity at the tree level. 
Consequently, the electroweak sector of the model is described by the four input parameters such as 
$\alpha_{\text{em}}$, $G_F$, $m_Z$ and $\sin^2\theta_W$. 
We calculate the one loop contribution to the W boson mass as well as to the rho parameter 
in order to clarify the possible mass spectrum of the extra Higgs bosons under the constraint from the electroweak precision data. 
We find that the hierarchical mass spectrum among $H^{\pm\pm}$, $H^{\pm}$ and $A$ (or $H$) 
is favored by the precision data 
especially for the case of $m_A$ $(\simeq m_H)>m_{H^+}>m_{H^{++}}$, 
where 
$H^{\pm\pm}$, $H^{\pm}$, $A$ and $H$ are the doubly-charged, singly-charged, CP-odd and CP-even Higgs bosons mainly 
originated from the triplet field. 
We also discuss phenomenological consequences of such a mass spectrum with 
relatively large mass splitting. 
The decay rate of the Higgs boson decay into two photons is 
evaluated under the constraint from the electroweak precision data, 
regarding the recent Higgs boson searches at the CERN LHC.

\pacs{\, 12.60.Fr, 12.15.Lk, 14.80.Cp}
\end{abstract}

\maketitle


\section{Introduction}

The Higgs boson search is underway at the LHC. 
The mass of the Higgs boson in the Standard Model (SM) has 
already been constrained to be between 115 GeV and 127 GeV or 
to be higher than 600 GeV at the 95\% confidence level by the recent direct search results~\cite{LHC-Higgs}. 
By the combination with the electroweak precision measurement at the LEP~\cite{LEP} and the SLC~\cite{SLC}, 
we may expect that a light Higgs boson exists as long as the Higgs boson interactions 
are of SM-like and that it will be discovered in near future. 

The SM is expected to be replaced by a new model at higher energies above the TeV scale, 
which may give reasonable explanations for 
tiny neutrino masses~\cite{neutrino-oscillation}, the origin of dark matter~\cite{DM} 
and the baryon asymmetry of the Universe~\cite{BAU}. 
The low energy effective theory of such a more fundamental model often contains 
a non-minimal Higgs sector. 
The structure of a Higgs sector strongly depends on the property of the corresponding new physics model at the high energy scale, 
so that experimental reconstruction of the Higgs sector is extremely important to determine 
new paradigm for physics beyond the SM. 

Once a Higgs boson is found at the LHC, its properties such as the mass, the width and the decay branching ratios 
will be thoroughly measured as precisely as possible. 
The precision measurement for Higgs boson properties can also be performed at future linear colliders such as the 
International Linear Collider (ILC) and the Compact Linear Collider (CLIC). 
Each extended Higgs model will be tested by preparing the full set of theoretical predictions on the 
observables which will be precisely measured at experiments at these colliders. 

The electroweak rho parameter has been one of the most important experimental hints to select 
possible structures of extended Higgs sectors. 
Its experimental value ($\rho\simeq 1$) would indicate that the Higgs sector is composed 
of one or more doublet fields (or with singlet fields). 
Such models naturally predict $\rho=1$ at the tree level 
because of the custodial symmetry in the kinetic term of their Higgs sectors. 
Radiative corrections give the deviation from unity corresponding to the violation of the custodial symmetry 
in the physics of dynamics in the loop~\cite{fermion_loop,Toussaint,Bertolini,Peskin_Wells,Osland,Taniguchi}. 
On the other hand, 
as another possibility, we may consider with some physics motivations 
the other class of   
Higgs models which contain scalar fields of larger representations under the $SU(2)_L$ gauge symmetry, 
in which $\rho\neq 1$ is predicted at the tree level. 
In these models, the vacuum expectation values (VEVs) of such fields with large isospin representations 
are highly constrained to be much smaller than 246 GeV to satisfy $\rho \simeq 1$. 

The Higgs triplet model (HTM) is one of the examples which predict $\rho \neq 1$ at the tree level.  
In the HTM, a Higgs triplet field with the hypercharge $Y=1$ 
is added to the minimal model with a Higgs doublet field with $Y=1/2$. 
The HTM can explain the tiny neutrino masses in a simple way 
via the so-called type II seesaw mechanism~\cite{typeII}. 
An interesting feature of the HTM is the existence of doubly-charged Higgs bosons in addition to 
singly-charged as well as extra neutral bosons, all of which are components of the triplet field. 

In this paper, we study radiative corrections to the electroweak observables in the HTM. 
In the model with $\rho \neq 1$ at the tree level like the HTM, 
apart from the models with only doublet fields and singlets which predict $\rho=1$ at the tree level, 
a new input parameter has to be introduced 
in addition to the usual three input parameters such as ($\alpha_{\text{em}}$, $G_F$ and $m_Z$).   
In Ref.~\cite{blank_hollik}, the on-shell renormalization scheme is constructed in the Higgs model with the $Y=0$ triplet field, in which 
four input electroweak parameters ($\alpha_{\text{em}}$, $G_F$, $m_Z$ and $\sin^2\theta_W$) are chosen to 
describe all the other electroweak observables. 
The radiative corrections to the electroweak observables have been calculated 
in the $Y=0$ triplet model~\cite{blank_hollik,real-triplet,Chen:2005jx} and in the left-right symmetric model~\cite{Chen:2005jx}.  

In our analysis, we first define the on-shell renormalization scheme for the electroweak sector of 
the HTM by using the method in Ref.~\cite{blank_hollik}. We then 
calculate radiative corrections to the electroweak observables such as $m_W$ and $\rho$ as 
a function of the four input parameters ($\alpha_{\text{em}}$, $G_F$, $m_Z$ and $\sin^2\theta_W$)\footnote{
In Ref.~\cite{Melfo:2011nx}, the constraint from the electroweak precision data has been discussed in the $Y=1$ HTM. 
However, the renormalization scheme they used seems to be only valid in models with $\rho=1$ at the tree level. 
}. 
We examine the preferable values of the mass spectrum 
of the triplet-like Higgs bosons and the VEV of the triplet field under the constraint from the electroweak precision data. 
We find that the hierarchical mass spectrum with large mass splitting is favored especially for the case of 
$m_A$ $(\simeq m_H)>m_{H^+}>m_{H^{++}}$, where 
$m_A$, $m_H$, $m_{H^+}$ and $m_{H^{++}}$ are the masses of the CP-odd ($A$), CP-even ($H$), singly-charged ($H^\pm$) and 
doubly-charged ($H^{\pm\pm}$) Higgs bosons, respectively, which mainly come from the components of the triplet field. 
On the contrary, the inverted hierarchical case with $m_{H^{++}}>m_{H^+}>m_A$ $(\simeq m_H)$ is 
relatively disfavored, in particular for the light SM-like Higgs boson. 
We then discuss implication with the Higgs boson phenomenology at the LHC 
in the HTM~\cite{gunion,delm01,delm02,4-lepton,TaoHan,Kadastik:2007yd,Akeroyd-Sugiyama,aky,Akeroyd_Moretti,Akeroyd_Moretti_Sugiyama}. 
The deviation due to the quantum effect of doubly- and singly-charged Higgs bosons is evaluated on the 
decay rate of $h\to \gamma\gamma$~\cite{Arhrib_hgg}, where $h$ is the SM-like Higgs boson. 
We here examine the decay rate in the parameter regions which are indicated by the electroweak precision data 
for the mass of the SM-like Higgs boson to be 125 GeV taking into account the recent Higgs boson search at the LHC~\cite{LHC-Higgs}. 
The deviation from the SM prediction can be significant although 
it is destructive. 
We also give some comments on the direct searches for the triplet-like Higgs bosons in the case 
with~\cite{Melfo:2011nx,Akeroyd-Sugiyama,aky,Akeroyd_Moretti,Akeroyd_Moretti_Sugiyama} 
and without~\cite{gunion,delm01,4-lepton,delm02,Kadastik:2007yd,TaoHan} the mass splitting among these Higgs bosons.  

In Sec.~II, we give a brief review for the tree level formulae in the HTM towards the one-loop 
calculation.  
In Sec.~III, the on-shell renormalization scheme is introduced for the electroweak precision observables, and 
the predictions for $m_W$ and $\rho$ at the one-loop level are calculated. 
In Sec.~IV, we evaluate the deviation from the SM prediction of the decay rate of $h\to \gamma\gamma$, and 
we discuss the collider phenomenology of the triplet-like Higgs bosons at the LHC. 
Discussions and conclusions are given in Sec.~V. 

\section{The Higgs Triplet Model}
The scalar sector of the HTM is composed of the isospin doublet field $\Phi$ with 
hypercharge $Y=1/2$ and the triplet field $\Delta$ with $Y=1$. 
The kinetic terms and relevant terms in the Lagrangian are given by 
\begin{align}
\mathcal{L}_{\text{HTM}}=(D_\mu \Phi)^\dagger (D^\mu \Phi)+\text{Tr}[(D_\mu \Delta)^\dagger (D^\mu \Delta)]
+\mathcal{L}_{Y}-V(\Phi,\Delta), 
\end{align}
where $\mathcal{L}_{Y}$ and $V(\Phi,\Delta)$ are the Yukawa interaction for neutrinos and the Higgs potential. 
From the kinetic term, 
the masses of the W boson and the Z boson are obtained at the tree level as 
\begin{align}
m_W^2 = \frac{g^2}{4}(v_\Phi^2+2v_\Delta^2),\quad m_Z^2 =\frac{g^2}{4\cos^2\theta_W}(v_\Phi^2+4v_\Delta^2), 
\end{align}
where $v_\Phi$ and $v_\Delta$ 
are the VEVs of the doublet Higgs field and the triplet Higgs field, respectively which satisfy 
$v^2\equiv v_\Phi^2+2v_\Delta^2\simeq$ (246 GeV)$^2$. 
The electroweak rho parameter can deviate from unity at the tree level; 
\begin{align}
\rho \equiv \frac{m_W^2}{m_Z^2\cos^2\theta_W}=\frac{1+\frac{2v_\Delta^2}{v_\Phi^2}}{1+\frac{4v_\Delta^2}{v_\Phi^2}}. \label{rho}
\end{align}
As the experimental value of the rho parameter is near unity, 
$v_\Delta^2/v_\Phi^2$ is required to be much smaller than unity.  

The Yukawa interaction for neutrinos is given by
\begin{align}
\mathcal{L}_Y&=h_{ij}\overline{L_L^{ic}}i\tau_2\Delta L_L^j+\text{h.c.}, 
\end{align}
where $L_L^i$ is the $i$-th generation left-handed lepton doublet, $h_{ij}$ is the $3\times 3$ complex symmetric Yukawa matrix. 
Notice that the triplet field $\Delta$ carries the lepton number of 2. 
The Majorana masses of neutrinos are generated by the Yukawa interaction with the VEV of the triplet field as
\begin{align}
(m_\nu)_{ij}=\sqrt{2}h_{ij} v_\Delta. 
\end{align}

The most general form of the Higgs potential under the gauge symmetry is given by 
\begin{align}
V(\Phi,\Delta)&=m^2\Phi^\dagger\Phi+M^2\text{Tr}(\Delta^\dagger\Delta)+\left[\mu \Phi^Ti\tau_2\Delta^\dagger \Phi+\text{h.c.}\right]\notag\\
&+\lambda_1(\Phi^\dagger\Phi)^2+\lambda_2\left[\text{Tr}(\Delta^\dagger\Delta)\right]^2+\lambda_3\text{Tr}[(\Delta^\dagger\Delta)^2]
+\lambda_4(\Phi^\dagger\Phi)\text{Tr}(\Delta^\dagger\Delta)+\lambda_5\Phi^\dagger\Delta\Delta^\dagger\Phi, 
\end{align}
where $m$ and $M$ are the dimension full real parameters, $\mu$ is the dimension full complex parameter 
which violates the lepton number, and 
$\lambda_1$-$\lambda_5$ are the coupling constants which are real. 
We here take $\mu$ to be real. 
The scalar fields $\Phi$ and $\Delta$ can be parameterized as
\begin{align}
\Phi=\left[
\begin{array}{c}
\varphi^+\\
\frac{1}{\sqrt{2}}(\varphi+v_\Phi+i\chi)
\end{array}\right],\quad \Delta =
\left(
\begin{array}{cc}
\frac{\Delta^+}{\sqrt{2}} & \Delta^{++}\\
\Delta^0 & -\frac{\Delta^+}{\sqrt{2}} 
\end{array}\right)\text{ with } \Delta^0=\frac{1}{\sqrt{2}}(\delta+v_\Delta+i\xi),
\end{align}
From the stationary condition at the vacuum $(v_\Phi, v_\Delta)$, we obtain 
\begin{align}
m^2&=\frac{1}{2}\left[-2v_\Phi^2\lambda_1-v_\Delta^2(\lambda_4+\lambda_5)+2\sqrt{2}\mu v_\Delta\right],\\
M^2&=M_\Delta^2-\frac{1}{2}\left[2v_\Delta^2(\lambda_2+\lambda_3)+v_\Phi^2(\lambda_4+\lambda_5)\right], \label{vc}
\text{ with } M_\Delta^2\equiv \frac{v_\Phi^2\mu}{\sqrt{2}v_\Delta}.
\end{align}
The mass matrices for the scalar bosons can be diagonalized by rotating the 
scalar fields as 
\begin{align}
\left(
\begin{array}{c}
\varphi^\pm\\
\Delta^\pm
\end{array}\right)&=
\left(
\begin{array}{cc}
c_{\beta_\pm} & -s_{\beta_\pm} \\
s_{\beta_\pm}   & c_{\beta_\pm}
\end{array}
\right)
\left(
\begin{array}{c}
w^\pm\\
H^\pm
\end{array}\right),\quad 
\left(
\begin{array}{c}
\chi\\
\xi
\end{array}\right)=
\left(
\begin{array}{cc}
c_{\beta_0} & -s_{\beta_0}\\
s_{\beta_0} & c_{\beta_0} 
\end{array}
\right)
\left(
\begin{array}{c}
z\\
A
\end{array}\right),\quad
\left(
\begin{array}{c}
\varphi\\
\delta
\end{array}\right)=
\left(
\begin{array}{cc}
c_\alpha & -s_\alpha\\
s_\alpha & c_\alpha
\end{array}
\right)
\left(
\begin{array}{c}
h\\
H
\end{array}\right),
\end{align}
with the mixing angles
\begin{align}
\tan\beta_\pm=\frac{\sqrt{2}v_\Delta}{v_\Phi},\quad \tan\beta_0 = \frac{2v_\Delta}{v_\Phi}, \quad
\tan2\alpha &=\frac{v_\Delta}{v_\Phi}\frac{2v_\Phi^2(\lambda_4+\lambda_5)-4M_\Delta^2}{2v_\Phi^2\lambda_1-M_\Delta^2-v_\Delta^2(\lambda_2+\lambda_3)}, \label{tan2a}
\end{align}
where we used abbreviation such as $s_\theta=\sin\theta$ and $c_\theta=\cos\theta$.  
In addition to the three Nambu-Goldstone bosons $w^\pm$ and $z$ which are absorbed by the longitudinal components 
of the $W$ boson and the $Z$ boson, 
there are seven physical mass eigenstates $H^{\pm\pm}$, $H^\pm$, $A$, $H$ and $h$. 
For the case of $v_\Delta^2/v_\Phi^2\ll 1$ which is required by the electroweak precision data, 
the state $h$ behaves mostly as the SM Higgs boson, while 
the other states are almost originated from the components of the triplet field. 
In this case, there are interesting relations among the masses; 
\begin{align}
&m_{H^{++}}^2-m_{H^{+}}^2\simeq m_{H^+}^2-m_A^2\left(\simeq-\frac{\lambda_5}{4}v_\Phi^2\equiv \xi\right), \label{mass1}\\
&m_H^2\simeq m_A^2\left(\simeq M_\Delta^2\right). \label{mass2}
\end{align}
These characteristic mass relations would be used as a probe of the Higgs potential in the HTM~\cite{aky}. 
If the masses of the triplet-like Higgs bosons are hierarchical, 
there are two patterns of the mass hierarchy among the triplet-like scalar bosons; i.e., 
when $\lambda_5$ is positive (negative), the mass hierarchy is $m_{\phi^0}>m_{H^+}>m_{H^{++}}$ ($m_{H^{++}}>m_{H^+}>m_{\phi^0}$), 
where $m_{\phi^0}=m_A$ or $m_H$. 
We here define the mass difference between $H^{\pm\pm}$ and $H^\pm$ as 
\begin{equation}
\Delta m \equiv m_{H^{++}}-m_{H^+}. 
\end{equation}

\section{One-loop corrections to electroweak parameters}

In this section, we calculate one-loop corrected electroweak observables   
in the on-shell scheme which was at first proposed by Blank and Hollik~\cite{blank_hollik} in the model with 
a triplet Higgs field with $Y=0$. 
In the SM, and in all the models with $\rho=1$ at the tree level, 
the kinetic term of the Higgs field contains three parameters $g$, $g'$ and $v$. 
All the electroweak parameters are determined by giving a set of three input parameters which 
are well known; i.e., for example, $\alpha_{\text{em}}$, $G_F$ and $m_Z$~\cite{hollik_sm,Aoki:1982ed}. 
On the other hand, in models with $\rho\neq 1$ at the tree level like the HTM, 
an additional input parameter is necessary to describe electroweak parameters. 
Therefore, in addition to the three input parameters $\alpha_{\text{em}}$, $G_F$ and $m_Z$, 
we take the weak angle $\sin^2\theta_W$ as the fourth input parameter in our calculation as in Ref.~\cite{blank_hollik}. 
The experimental values of these input parameters are given by~\cite{pdg}
\begin{align}
&\alpha_{\text{em}}^{-1}(m_Z) = 128.903(15),\quad G_F= 1.16637(1)\times 10^{-5} \text{ GeV}^{-2},\notag\\
&m_Z=91.1876(21)\text{ GeV},\quad \hat{s}_W^2(m_Z)= 0.23146(12),  \label{data}
\end{align}
where $\hat{s}_W^2(m_Z)$ is defined as the ratio of the coefficients of the vector part and the axial vector part in the 
$Z\bar{e}e$ vertex;
\begin{align}
1-4\hat{s}_W^2(m_Z)=\frac{\text{Re}(v_e)}{\text{Re}(a_e)}, 
\end{align}
where $v_e$ and $a_e$ are defined in Eq.~(\ref{ren_zee}). 
Tree level formulae for the other electroweak parameters are given in terms of the four input parameters:
\begin{align}
g^2&=\frac{4\pi\alpha_{\text{em}}}{\hat{s}_W^2},\label{gsq}\\
m_W^2&=\frac{\pi\alpha_{\text{em}}}{\sqrt{2}G_F\hat{s}_W^2},\label{mwsq}\\
v^2&=\frac{1}{\sqrt{2}G_F},\label{v} \\
v_\Delta^2 &= \frac{\hat{s}_W^2\hat{c}_W^2}{2\pi\alpha_{\text{em}}}m_Z^2-\frac{\sqrt{2}}{4G_F}\label{vdel}, 
\end{align}
where $\hat{s}_W^2=\hat{s}_W^2(m_Z)$ and $\hat{c}_W^2=1-\hat{s}_W^2$
\footnote{In the limit of $v_\Delta\to 0$, we obtain the following relation
\begin{align}
\frac{\hat{s}_W^2\hat{c}_W^2}{\pi\alpha_{\text{em}}}m_Z^2=\frac{1}{\sqrt{2}G_F}.
\end{align}
By using this relation, $m_W$ can be expressed by 
$m_W^2 =m_Z^2\hat{c}^2_W$. 
This relation can be found in models with $\rho=1$ at the tree level such as the SM. }.
The deviation from the relation in Eq.~(\ref{mwsq}) due to radiative corrections can be parameterized as 
\begin{align}
G_F = \frac{\pi \alpha_{\text{em}}}{\sqrt{2}m_W^2\hat{s}_W^2}(1+\Delta r), \label{gf}
\end{align}
where $\Delta r$ is obtained as~\cite{blank_hollik}
\begin{align}
\Delta r &= \frac{\Pi_T^{WW}(0)-\Pi_T^{WW}(m_W^2)}{m_W^2}+\frac{d}{dp^2}\Pi_T^{\gamma\gamma}(p^2)\Big|_{p^2=0}
+\frac{2\hat{s}_W}{\hat{c}_W}\frac{\Pi_T^{\gamma Z}(0)}{m_Z^2}-\frac{\hat{c}_W}{\hat{s}_W}
\frac{\Pi_T^{\gamma Z}(m_Z^2)}{m_Z^2}+\delta_{VB}+\delta_{V}'. \label{delta_r}
\end{align}
where $\delta_{VB}$ and $\delta_{V}'$ are the vertex and the box diagram corrections
to $G_F$ and the radiative corrections to the $Z\bar{e}e$ vertex, respectively.  
These are calculated as \cite{blank_hollik,hhkm} 
\begin{align}
\delta_{VB}  = \frac{\alpha_{\text{em}}}{4\pi \hat{s}_W^2}\left[6+\frac{10-10\hat{s}_W^2-3m_W^2/(\hat{c}_W^2m_Z^2)(1-2\hat{s}_W^2)}{2\left(1-\frac{m_W^2}{m_Z^2}\right)}\ln \frac{m_W^2}{m_Z^2}\right],\quad 
\delta_{V}'  = \frac{v_e}{2\hat{s}_W^2}\left[\frac{\Gamma_V^{Z\bar{e}e}(m_Z^2)}{v_e}-\frac{\Gamma_A^{Z\bar{e}e}(m_Z^2)}{a_e}\right], 
\end{align}
where $\Gamma_{V,A}^{Z\bar{e}e}$ are defined through the renormalized $Z\bar{e}e$ vertex as \cite{blank_hollik}
\begin{align}
\hat{\Gamma}_\mu^{Z\bar{e}e}(m_Z^2)&=i\frac{g}{2\hat{c}_W}\left[(v_e+a_e\gamma_5)\gamma_\mu+\gamma_\mu\Gamma_V^{Z\bar{e}e}(m_Z^2)
+\gamma_\mu\gamma_5\Gamma_A^{Z\bar{e}e}(m_Z^2)\right],\\
v_e&=-\frac{1}{2}+2\hat{s}_W^2,\quad a_e = -\frac{1}{2}. \label{ren_zee}
\end{align}
In Eq.~(\ref{delta_r}), $\Pi_T^{XY}(p^2)$ ($X,Y=W,Z,\gamma$) are the 1PI diagrams for the gauge boson self-energies. 
We show the list of all the analytic expressions for the gauge boson self-energies in the HTM with the $Y=1$ triplet field in Appendix. 
From Eqs.~(\ref{rho}) and (\ref{gf}), 
the renormalized W boson mass as well as the renormalized rho parameter are given by\footnote{
By keeping only the dependence of the top quark mass, $\Delta r$ can be written 
\begin{align}
\Delta r&\to \frac{g^2}{16\pi^2}\left[\frac{m_t^4}{2m_W^4}+\frac{m_t^2}{4m_W^2}-\frac{5}{3}
+\frac{1}{2}\left(\frac{m_t^6}{m_W^6}-3\frac{m_t^2}{m_W^2}\right)\ln\left(1-\frac{m_W^2}{m_t^2}\right)+\ln m_t^2
+\ln\left(1-\frac{m_W^2}{m_t^2}\right)\right]\notag\\
&-\frac{g^2s_W^2}{16\pi^2}4Q_t^2\ln m_t^2-\frac{g^2}{16\pi^2}(-4s_W^2Q_t^2+Q_t)\left(\ln m_t^2+\frac{1}{3}\right)\notag\\
&=
\frac{g^2}{16\pi^2}\left(-\frac{1}{3}+\ln m_t^2\right)-\frac{g^2}{16\pi^2}Q_t\left(\ln m_t^2+\frac{1}{3}\right)+\mathcal{O}(m_W^2/m_t^2), 
\end{align}
Thus, $\Delta r$ is depending on the top quark mass logarithmic. }
\begin{align}
m_W^2= \frac{\pi \alpha_{\text{em}}}{\sqrt{2}G_F\hat{s}_W^2}(1+\Delta r),\quad
\rho =  \frac{\pi \alpha_{\text{em}}}{\sqrt{2}G_Fm_Z^2\hat{s}_W^2\hat{c}_W^2}(1+\Delta r). 
\end{align}
Therefore, with four input parameters ($\alpha_{\text{em}}$, $G_F$, $m_Z$ and $\hat{s}_W^2$), 
$\Delta r$ determines both the one-loop corrected mass of the W boson and the rho parameter in the HTM. 

\begin{figure}[t]
\begin{center}
\includegraphics[width=85mm]{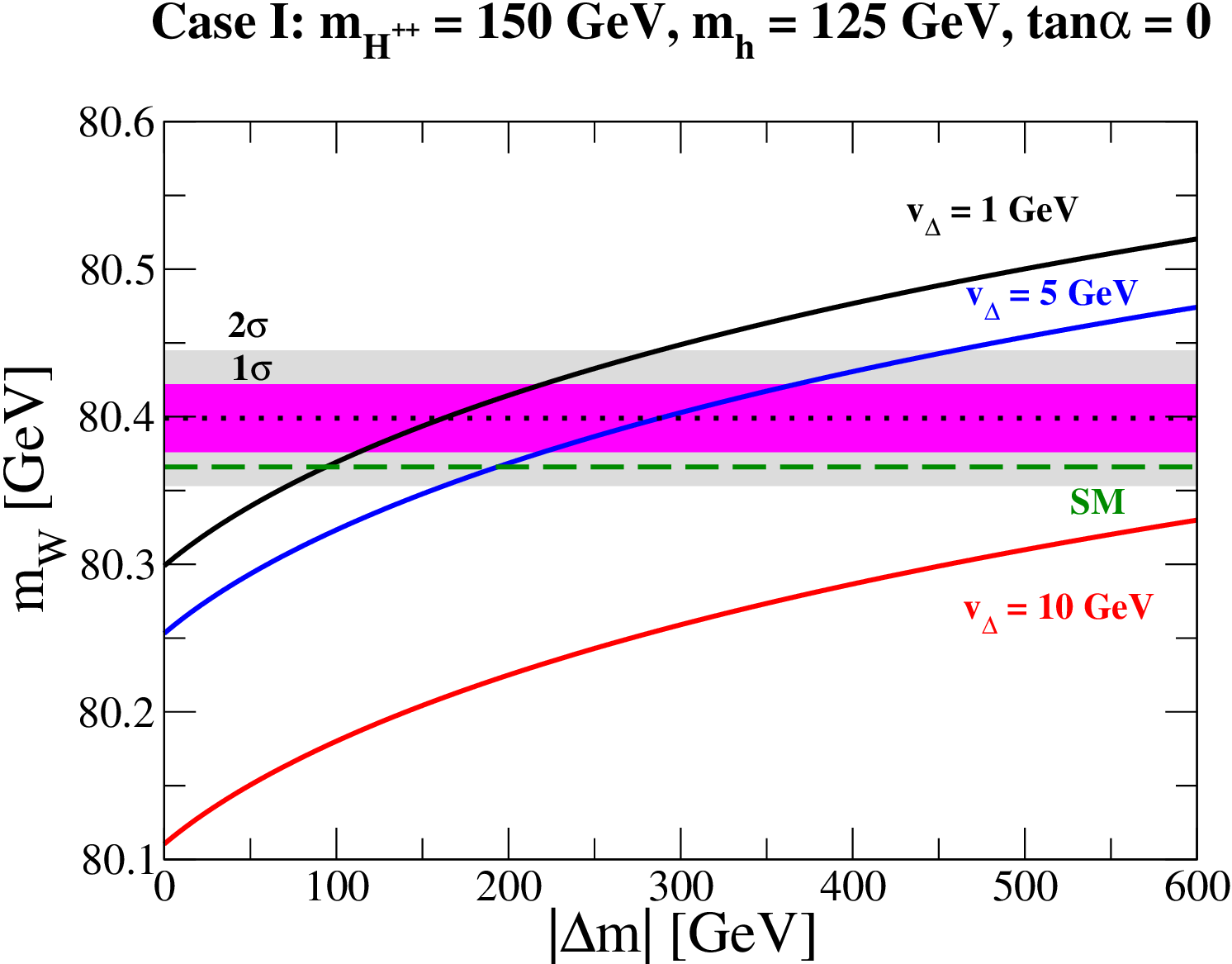}\hspace{3mm}
\includegraphics[width=85mm]{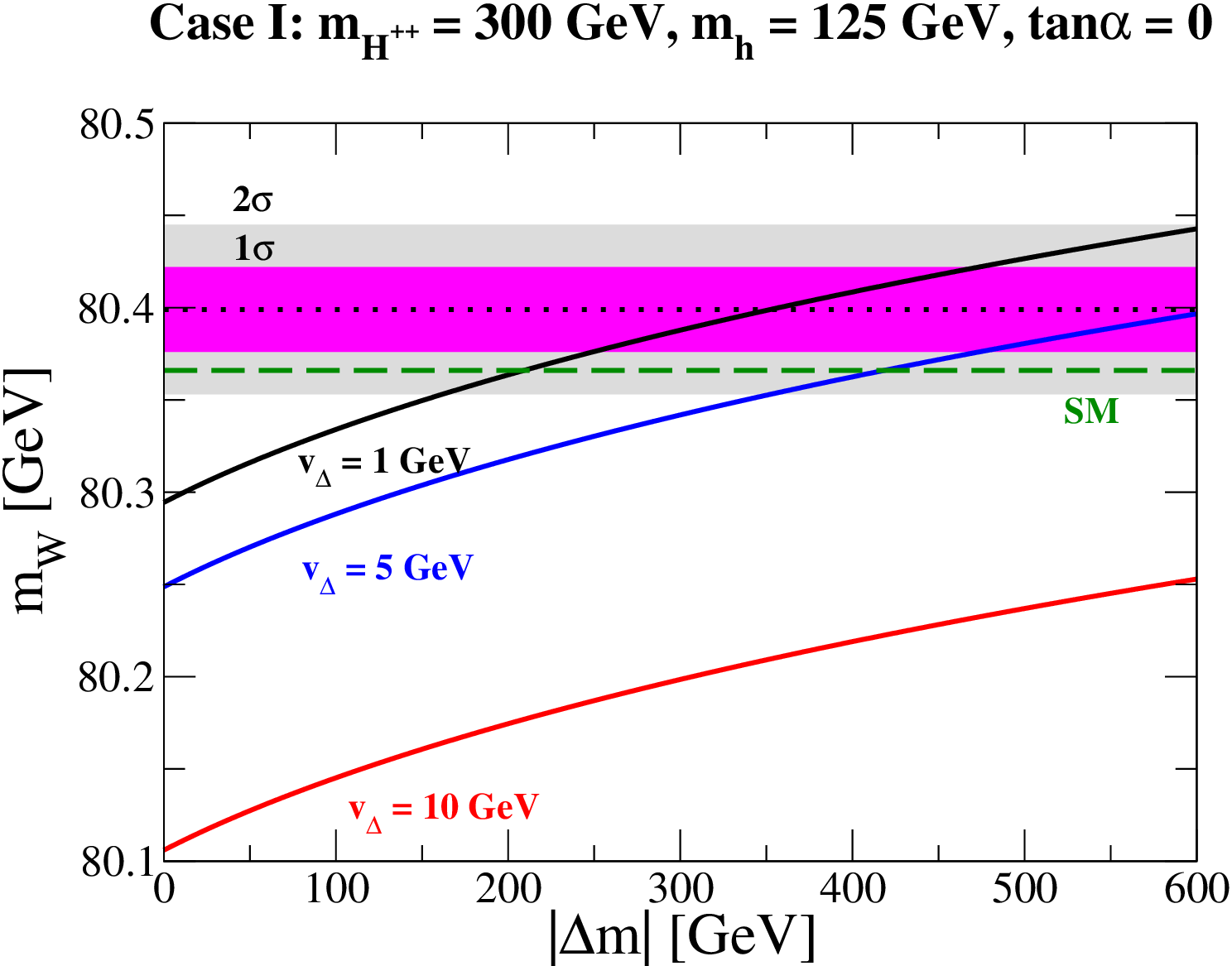}
\end{center}
\caption{The one-loop corrected values of $m_W$ as a function of the absolute value of $\Delta m$ 
for each fixed value of $v_\Delta$ in Case I ($m_{\phi^0} > m_{H^{+}}>m_{H^{++}}$). 
We take $m_h=125$ GeV, $m_t=173$ GeV and $\tan\alpha=0$ in the both figures. 
The pink (gray) shaded region represents the 1$\sigma$ (2$\sigma$) error for the experimental data of $m_W^{\text{exp}}$ 
(=80.399 $\pm$ 0.23 GeV~\cite{pdg}). 
In the left (right) figure, we take $m_{H^{++}}=$ 150 GeV (300 GeV). 
The dashed line shows the SM prediction of $m_W$ at the one-loop level with the SM Higgs boson mass to be 125 GeV. }
\label{mw1}
\end{figure}

In the following, we show numerical results for the radiative corrections to $m_W^2$ as well as $\rho$ 
in the HTM. 
The radiative correction depends on the mass spectrum and the mixing angle 
in the Higgs potential; i.e., $m_{H^{++}}$, $m_{H^{+}}$, $m_A$, $m_H$, $m_h$ and $\tan\alpha$.  
Although these six parameters are all free parameters, 
under the requirement of $v_\Delta^2\ll v^2$ the approximated formulae given in Eqs.~(\ref{mass1}) and (\ref{mass2})
tell us to pick up the following three parameters such as the mass of the SM-like Higgs boson $m_h$, 
the mass of the lightest triplet-like scalar boson $m_A$ (or $m_{H^{++}}$) and 
the mass difference $\Delta m$ between $H^{\pm\pm}$ and $H^\pm$. 
In the following analysis, we consider the scenarios with mass splitting for the triplet-like Higgs bosons; 
namely, for Case I ($m_{\phi^0} >m_{H^+}>m_{H^{++}}$) and Case II ($m_{H^{++}}>m_{H^+}>m_{\phi^0}$). 
We take pole masses of the top quark $m_t=173$ GeV and  
the bottom quark $m_b=4.7$ GeV and $\alpha_s(m_Z)=0.118$~\cite{pdg}. 
We take into account the leading order QCD correction in the calculation of the one-loop corrected $m_W$ as well as $\rho$.

In Fig.~\ref{mw1}, 
the renormalized value of $m_W$ is shown as a function of $|\Delta m|$ for several values of $v_\Delta$ 
in Case I with the data $m_W^{\text{exp}}=80.399\pm 0.023$ GeV~\cite{pdg}. 
The mass of the SM-like Higgs boson $h$ is taken as $m_h=125$ GeV, and the mixing angle $\tan\alpha$ is set on zero. 
The mass of the lightest triplet-like Higgs boson $m_{H^{++}}$ is taken to be 150 GeV (left figure) 
and 300 GeV (right figure). 
It is seen that the predicted value of $m_W$ for the degenerate mass case ($|\Delta m|=0$) is 
outside the region within the $2\sigma$ error. 
The prediction satisfies the data when $|\Delta m|$ has a non-zero value. 
When $m_{H^{++}}=150$ GeV, the favored value of $|\Delta m|$ by the data within the $2\sigma$ error 
is $80$ GeV$\lesssim |\Delta m|\lesssim 280$ GeV ($190$ GeV$\lesssim |\Delta m|\lesssim430$ GeV) for 
$v_\Delta=1$ GeV (5 GeV). 
The preferred value of $|\Delta m|$ for smaller values of $v_\Delta$ than 1 GeV is similar to that for $v_\Delta=1$ GeV. 
When $m_{H^{++}}=300$ GeV, the allowed values of $|\Delta m|$ are larger than the case of $m_{H^{++}}=150$ GeV for 
the same value of $v_\Delta$. 
Smaller mass splitting which satisfies the data corresponds to the smaller value of $v_\Delta$ while 
largest value of $|\Delta m|$ ($\sim$ 500-600 GeV), which comes from perturbative unitarity, corresponds to 
$v_\Delta\sim \mathcal{O}(10)$ GeV. 
We note that the result is almost unchanged even if we vary $\tan\alpha$ in the region of $0<\tan\alpha<1$ as long as 
$H^{\pm\pm}$ is not too heavy.  

\begin{figure}[t]
\begin{center}
\includegraphics[width=85mm]{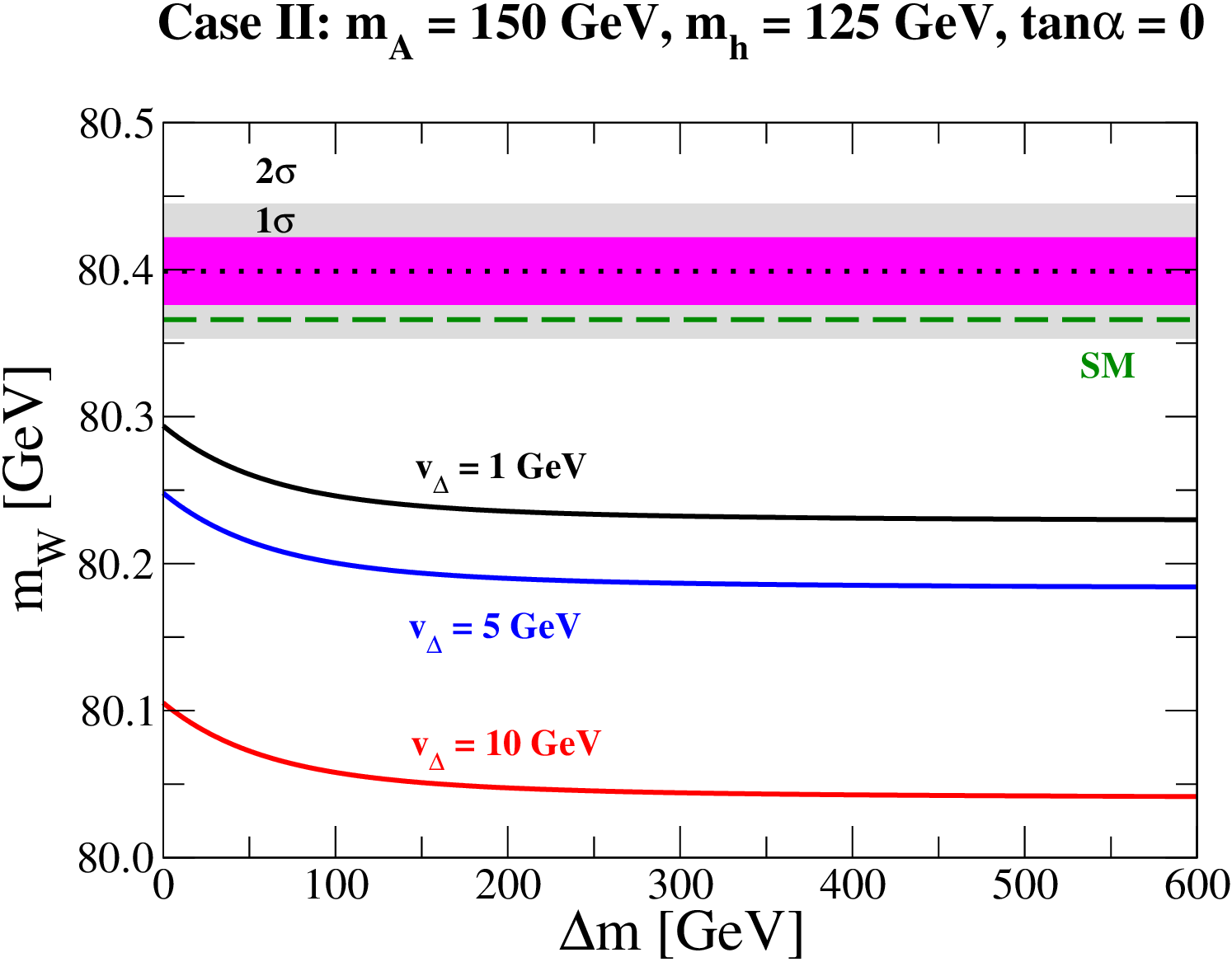}\hspace{3mm}
\includegraphics[width=85mm]{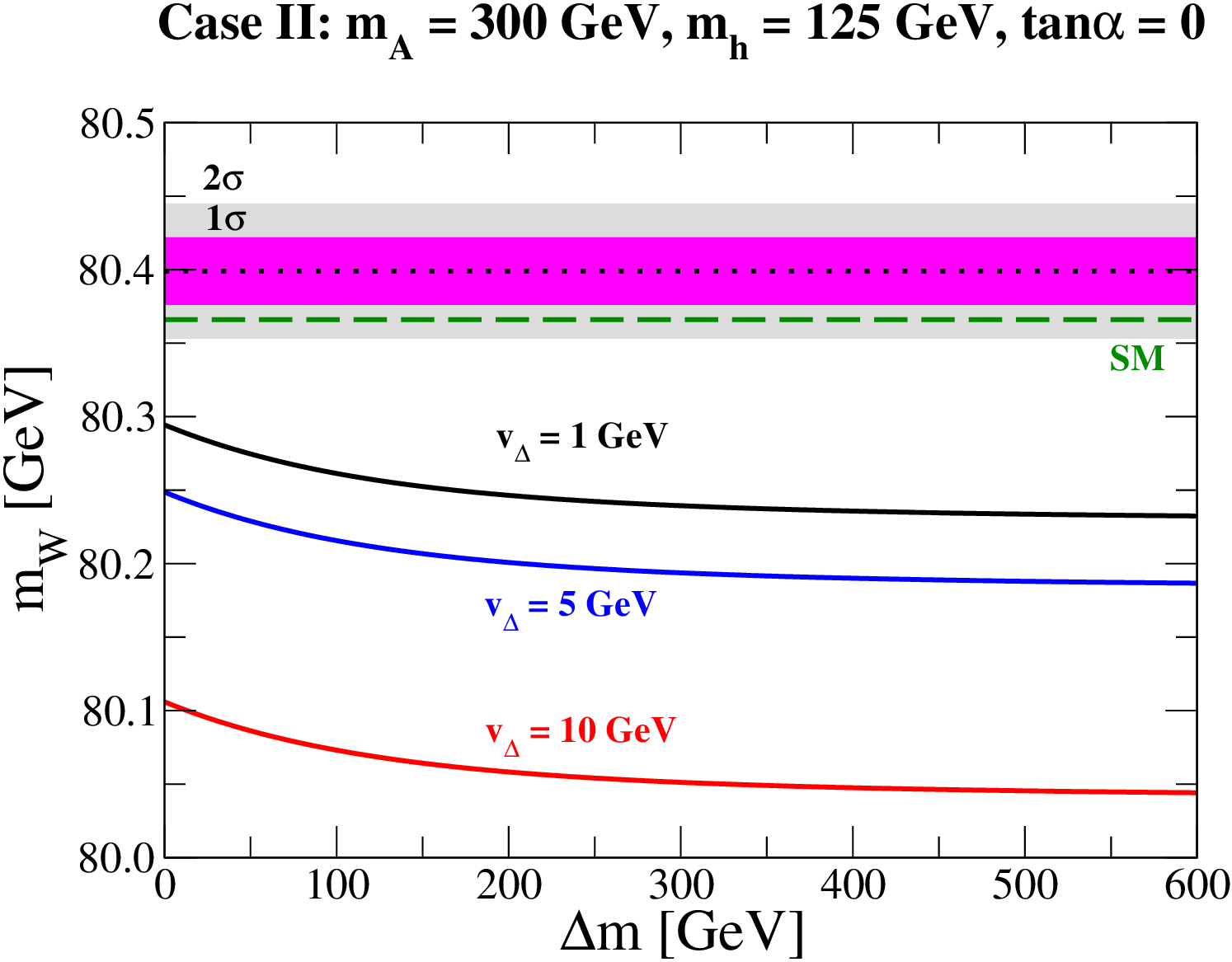}
\end{center}
\caption{The one-loop corrected values of $m_W$ as a function of $\Delta m$ 
for each fixed value of $v_\Delta$ in Case II ($m_{H^{++}}>m_{H^+}>m_{\phi^0}$). 
We take $m_h=125$ GeV, $m_t=173$ GeV and $\tan\alpha=0$ in the both figures. 
The pink (gray) shaded region represents the 1$\sigma$ (2$\sigma$) error for the experimental data of $m_W^{\text{exp}}$ 
(=80.399 $\pm$ 0.23 GeV~\cite{pdg}). 
In the left (right) figure, we take $m_{A}=$ 150 GeV (300 GeV). 
The dashed line shows the SM prediction of $m_W$ at the one-loop level with the SM Higgs boson mass to be 125 GeV. }
\label{mw2}
\end{figure}

In Fig.~\ref{mw2}, 
the renormalized value of $m_W$ is shown as a function of $\Delta m$ for several values of $v_\Delta$ 
in Case II with the data $m_W^{\text{exp}}=80.399\pm 0.023$ GeV~\cite{pdg}. 
The mass of the SM-like Higgs boson $h$ is taken as $m_h=125$ GeV, and $\tan\alpha$ is set on zero. 
The mass of the lightest triplet-like Higgs boson $m_A$ is taken to be 150 GeV (left figure) 
and 300 GeV (right figure). 
It is found that Case II is strongly constrained by the electroweak precision data for entire range of $\Delta m$. 
The situation is worse for larger values of $\Delta m$ and also for larger values of $v_\Delta$. 
Therefore, Case I can be more consistent with the electroweak precision data than the degenerate mass case and also Case II. 
We note that the result is almost unchanged even if we vary $\tan\alpha$ in the region of $0<\tan\alpha<1$ as long as 
$A$ is not too heavy. 

\begin{figure}[t]
\begin{center}
\includegraphics[width=85mm]{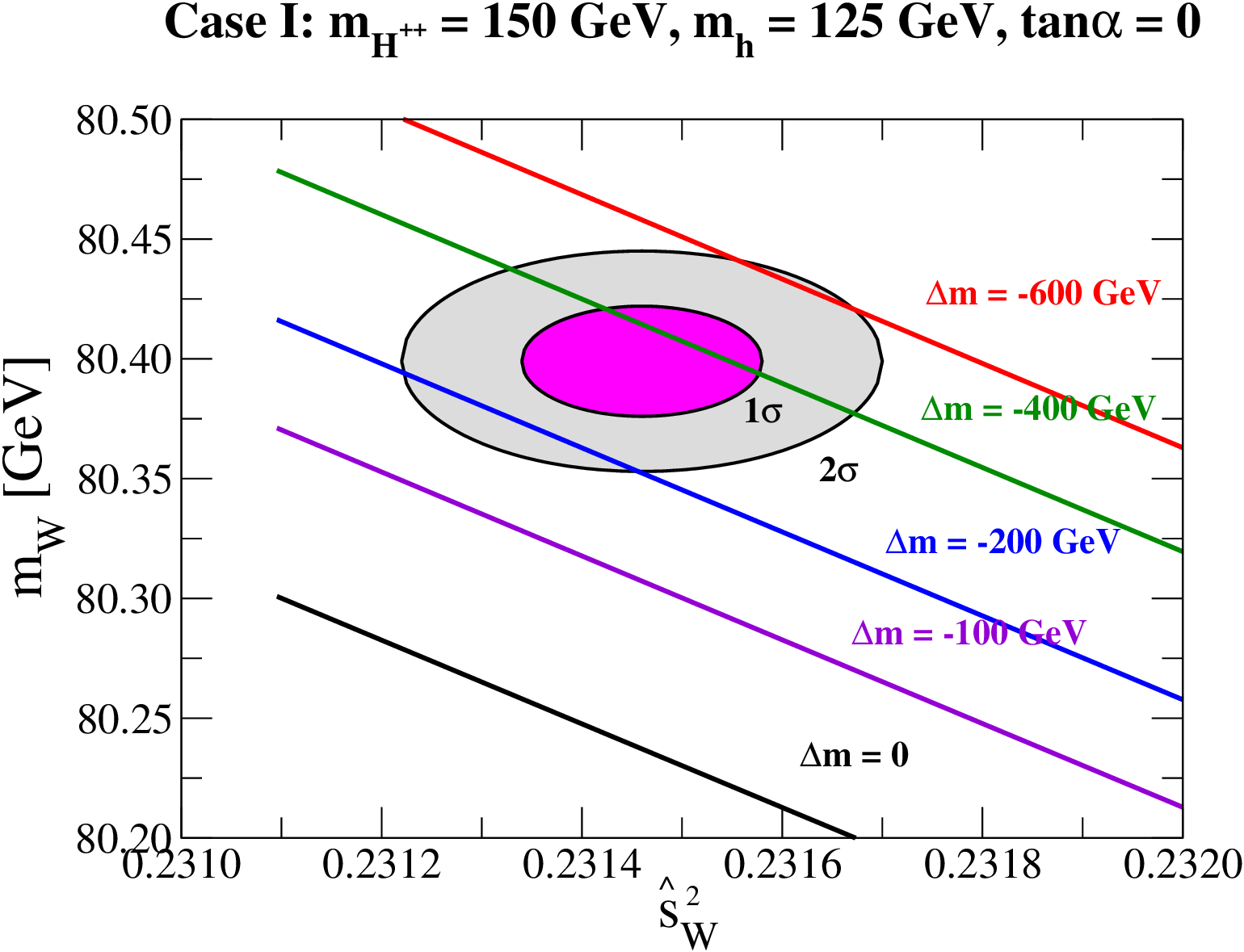}\hspace{3mm}
\includegraphics[width=85mm]{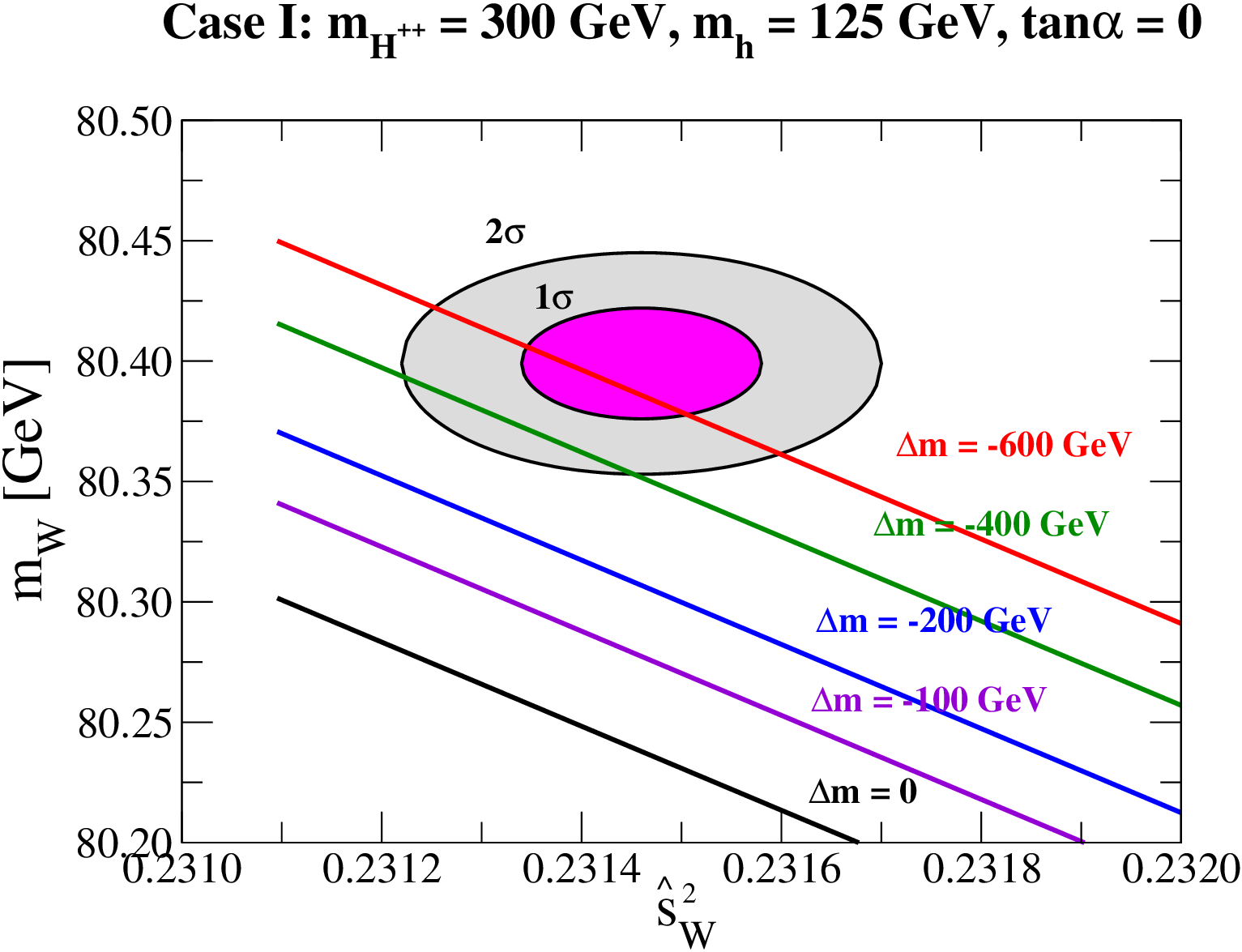}\\\vspace{6mm}
\includegraphics[width=85mm]{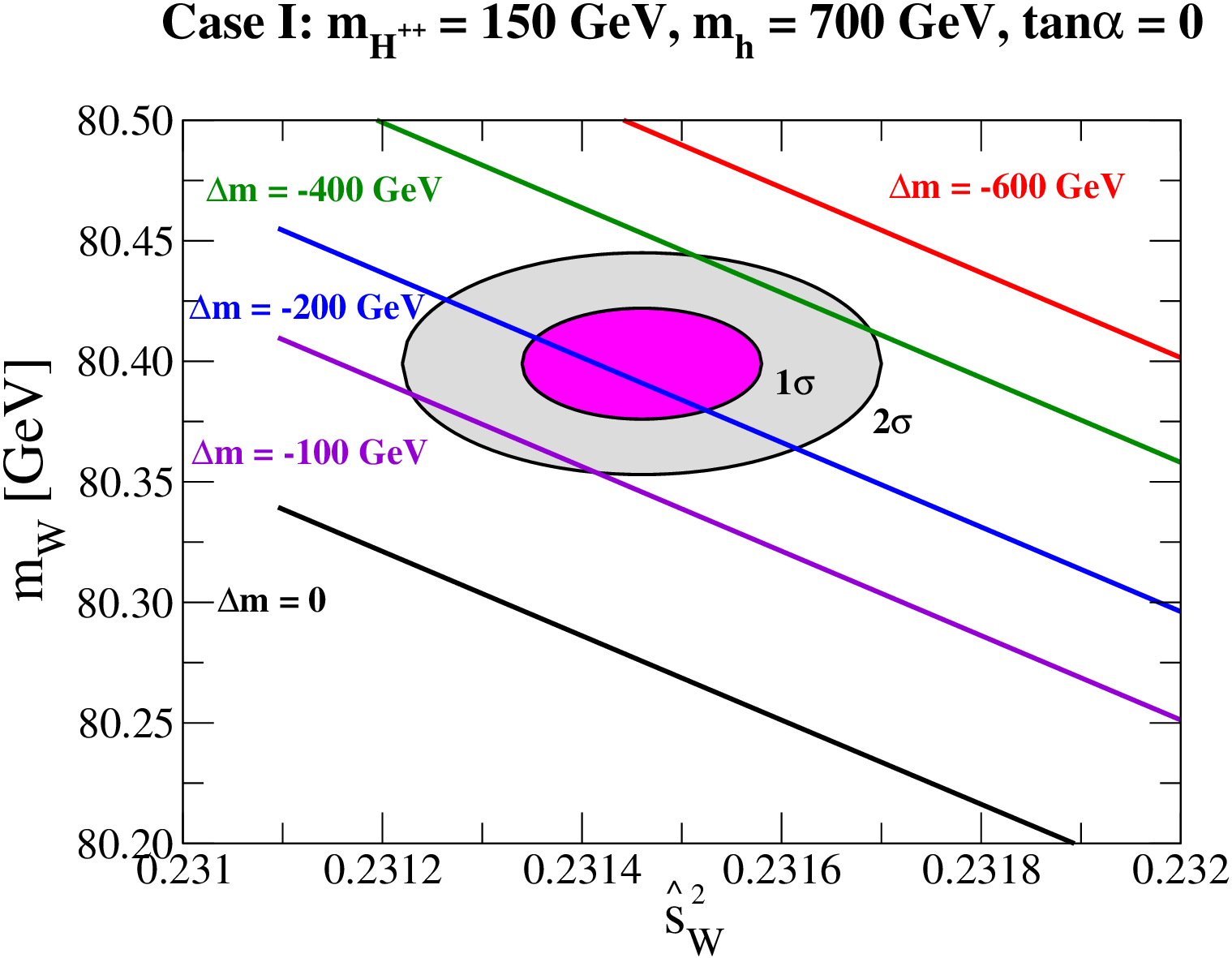}\hspace{3mm}
\includegraphics[width=85mm]{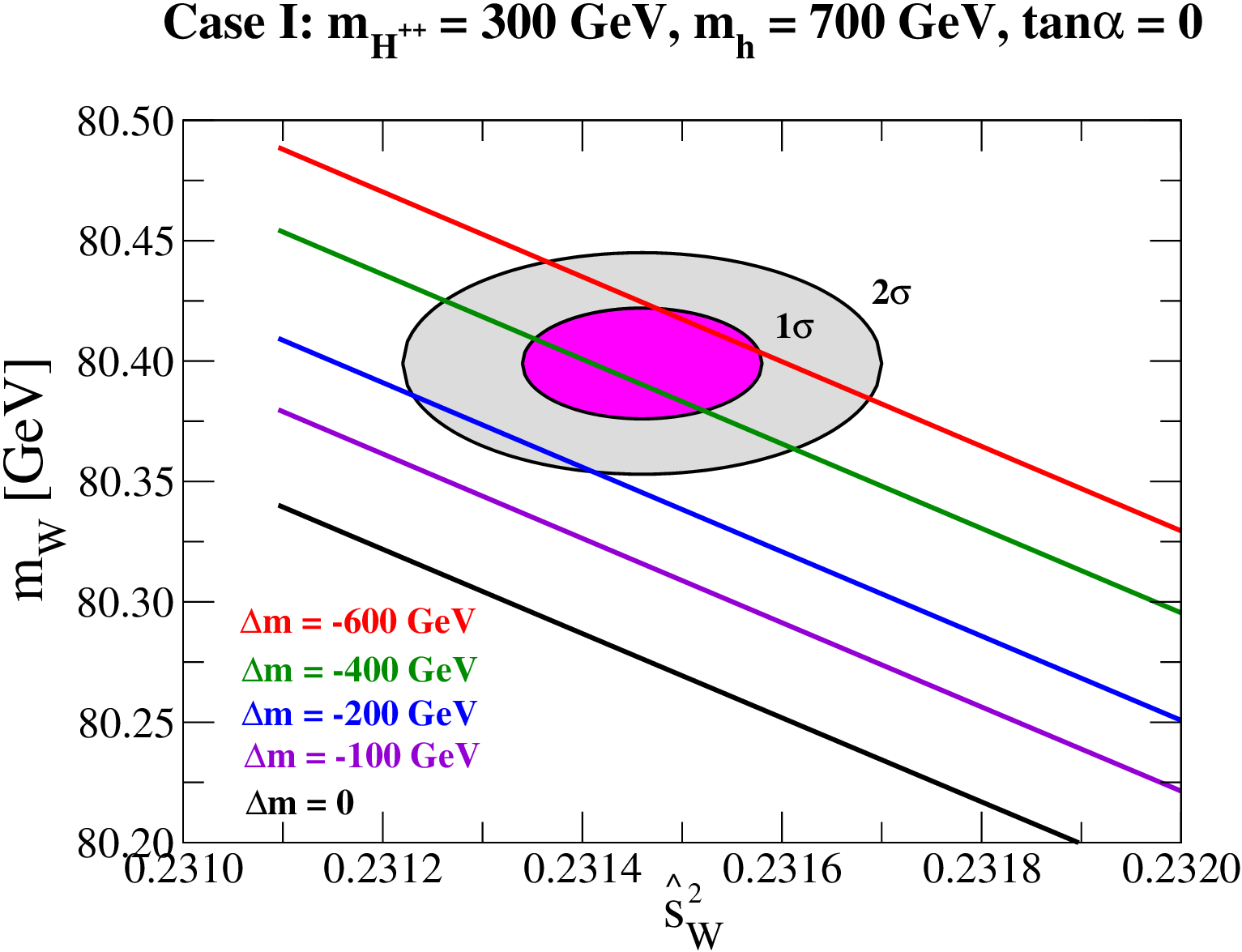}
\end{center}
\caption{The one-loop corrected values of $m_W$ as a function of $\hat{s}_W^2$ in Case I ($m_{\phi^0} > m_{H^{+}}>m_{H^{++}}$). 
We take $m_t=173$ GeV and $\tan\alpha=0$ in all the figures. 
The pink (gray) shaded region represents the 1$\sigma$ (2$\sigma$) error for the experimental data of 
$m_W^{\text{exp}}$ (=80.399 $\pm$ 0.23 GeV~\cite{pdg}) and $\hat{s}_W^{2\text{ exp}}$ (=0.23146 $\pm$ 0.00012 GeV~\cite{pdg}). 
In the left (right) two figures, we take $m_h=$ 125 GeV (700 GeV). 
The mass of the lightest triplet-like Higgs boson is taken to be 150 GeV and 300 GeV. }
\label{mw3}
\end{figure}

In Fig.~\ref{mw3}, 
we show the renormalized values for $m_W$ for each value of $\Delta m$ as a function of the input parameter $\hat{s}_W^2$ in Case I.  
The values of $(m_{H^{++}},m_h)$ are taken to be (150 GeV,125 GeV), (300 GeV,125 GeV), (150 GeV,700 GeV) and (300 GeV,700 GeV) 
in the figures located at the upper left, the upper right, the lower left and the lower right panels, respectively. 
In all figures, the mixing angle $\tan\alpha$ is set to be zero. 
Regions indicated by the data of $m_W$ and $\hat{s}_W^2$ within the $1\sigma$ error and the $2\sigma$ error are 
also shown in each figure for the fixed value of $m_t$ (=173 GeV). 
When $m_h=125$ GeV (upper figures), the predicted values of $m_W$ for $\Delta m =0$ are 
far from the allowed region by the data. 
For $m_{H^{++}}=150$ GeV and $m_h=125$ GeV (upper left), 
the prediction is consistent with the data within the $2\sigma$ error when about
$160$ GeV$\lesssim |\Delta m| \lesssim 600$ GeV is taken. 
On the other hand, for $m_{H^{++}}=300$ GeV and $m_h=125$ GeV (upper right), 
smaller values are predicted for $m_W$ as compared to the case with $m_{H^{++}}=150$ GeV for 
non-zero value of $|\Delta m|$. 
They approach to the predicted values of $m_W$ with $|\Delta m|=0$ in the large mass limit for $H^{\pm\pm}$. 
It is consistent with the data when we take $\Delta m\gtrsim 400$ GeV in this case. 
When $m_h=700$ GeV (lower figures), the predicted values of $m_W$ for $\Delta m =0$ are 
far from the allowed region by the data but closer than the case of $m_h=125$ GeV. 
For $m_{H^{++}}=150$ GeV and $m_h=700$ GeV (lower left), 
the prediction is consistent with the data within the $2\sigma$ error when about 
$100$ GeV$\lesssim |\Delta m| \lesssim 400$ GeV is taken. 
On the other hand, for $m_{H^{++}}=300$ GeV and $m_h=700$ GeV (lower right), 
it is consistent with the data when we take $\Delta m\gtrsim 200$ GeV in this case. 
The edge of each curve at $\hat{s}_W^2\simeq 0.2311$ corresponds to $v_\Delta=0$. 
We note that the result is almost unchanged even if we vary $\tan\alpha$ in the region of $0<\tan\alpha<1$ as long as 
$H^{\pm\pm}$ is not too heavy. 

\begin{figure}[t]
\begin{center}
\includegraphics[width=85mm]{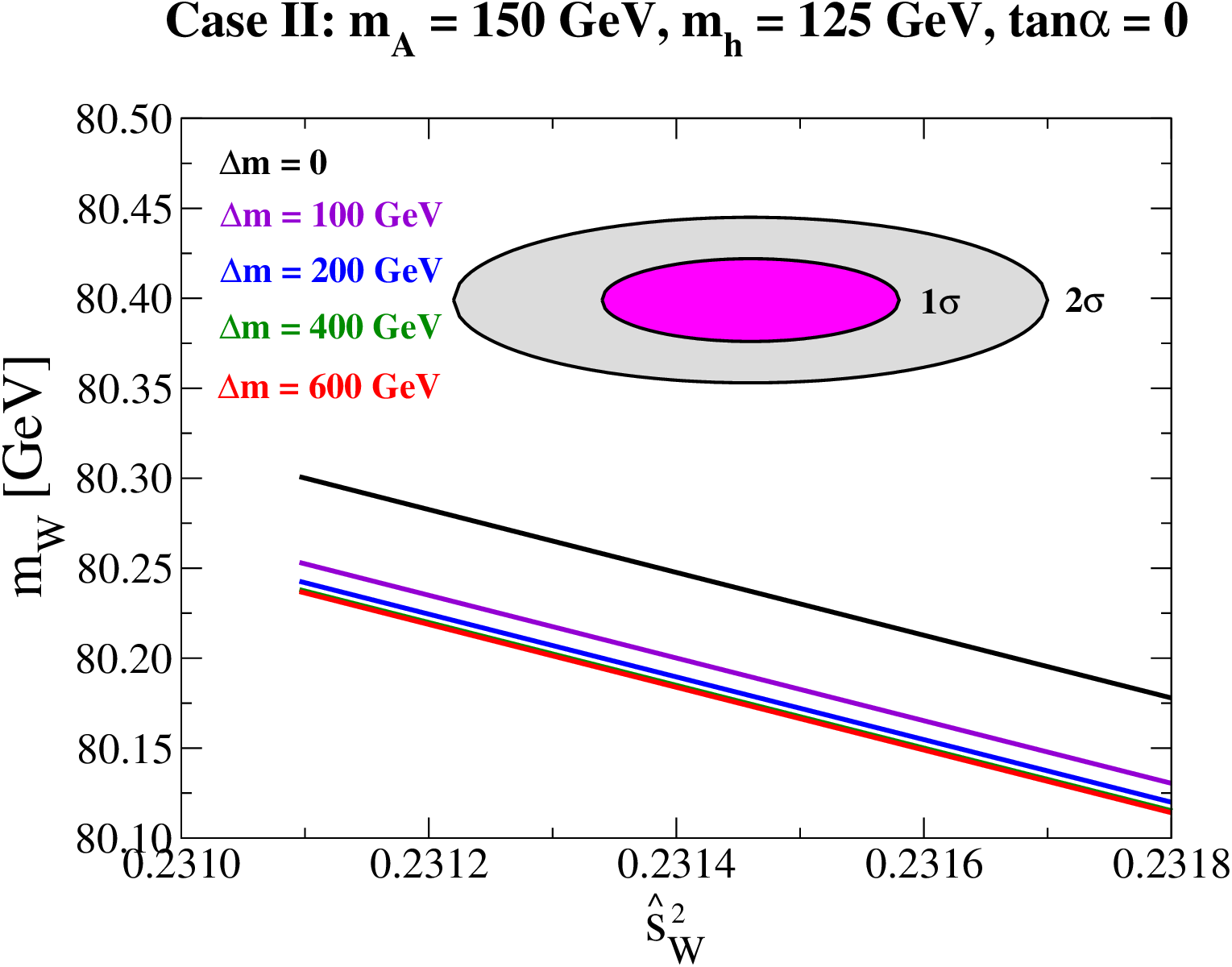}\hspace{3mm}
\includegraphics[width=85mm]{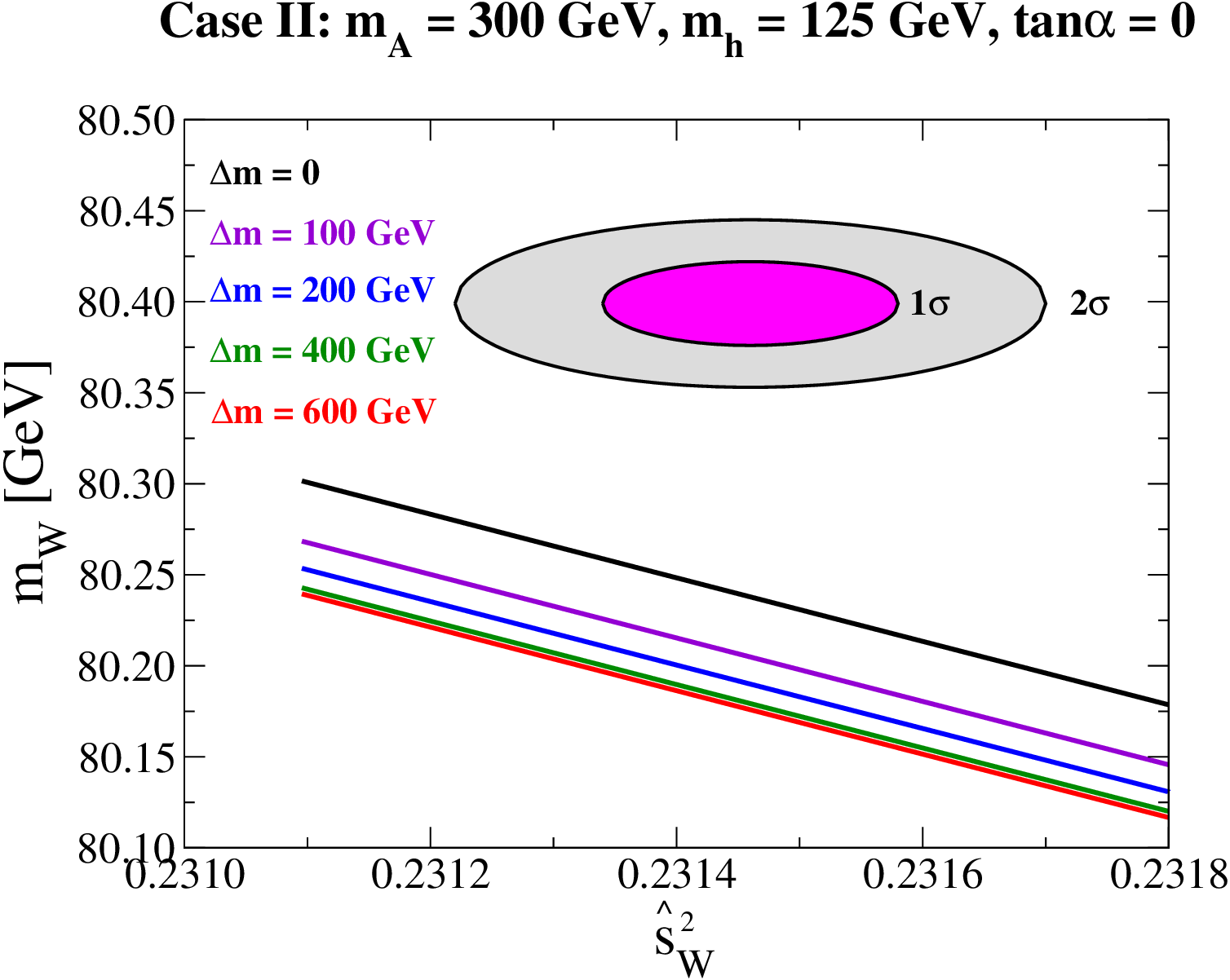}\\\vspace{6mm}
\includegraphics[width=85mm]{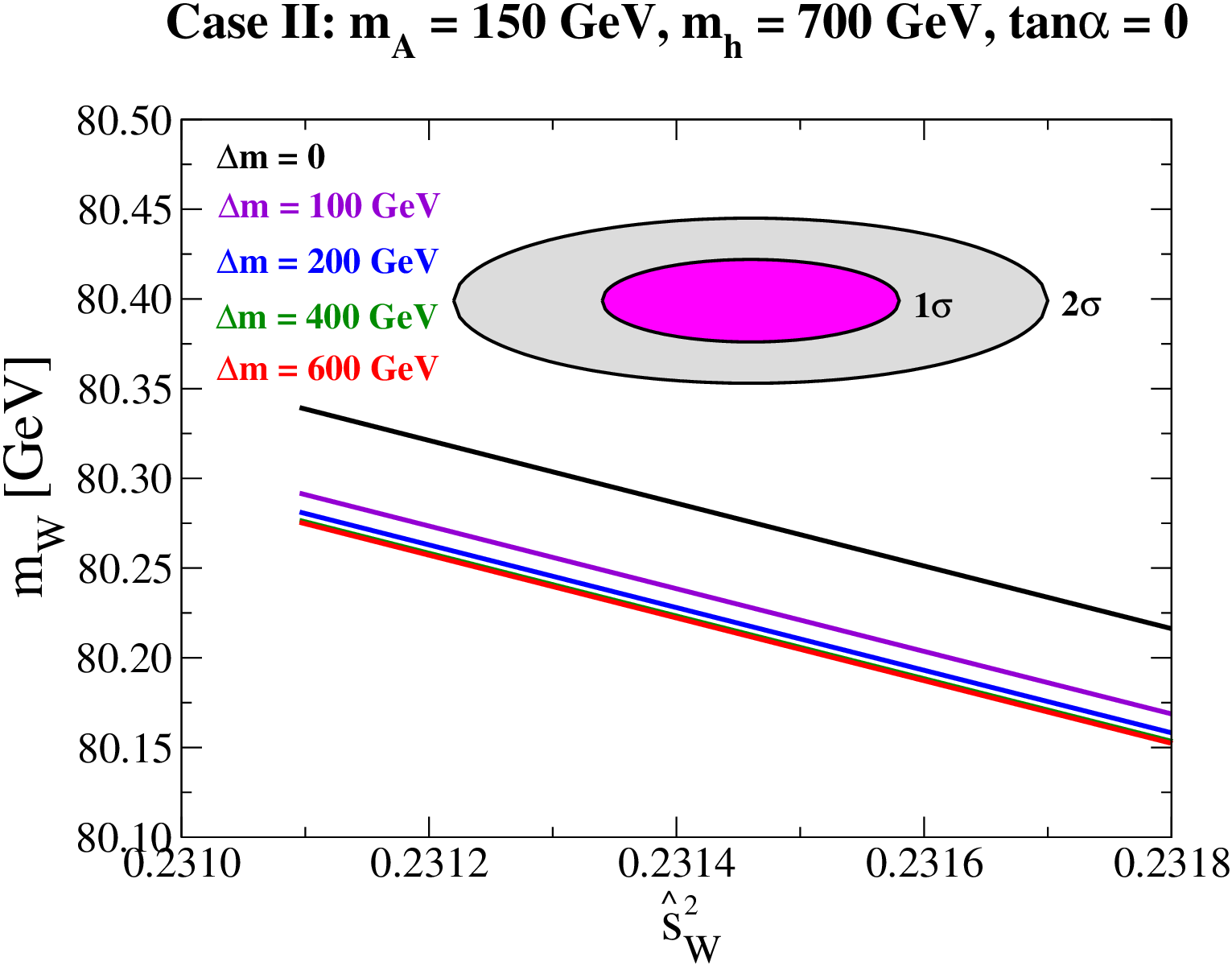}\hspace{3mm}
\includegraphics[width=85mm]{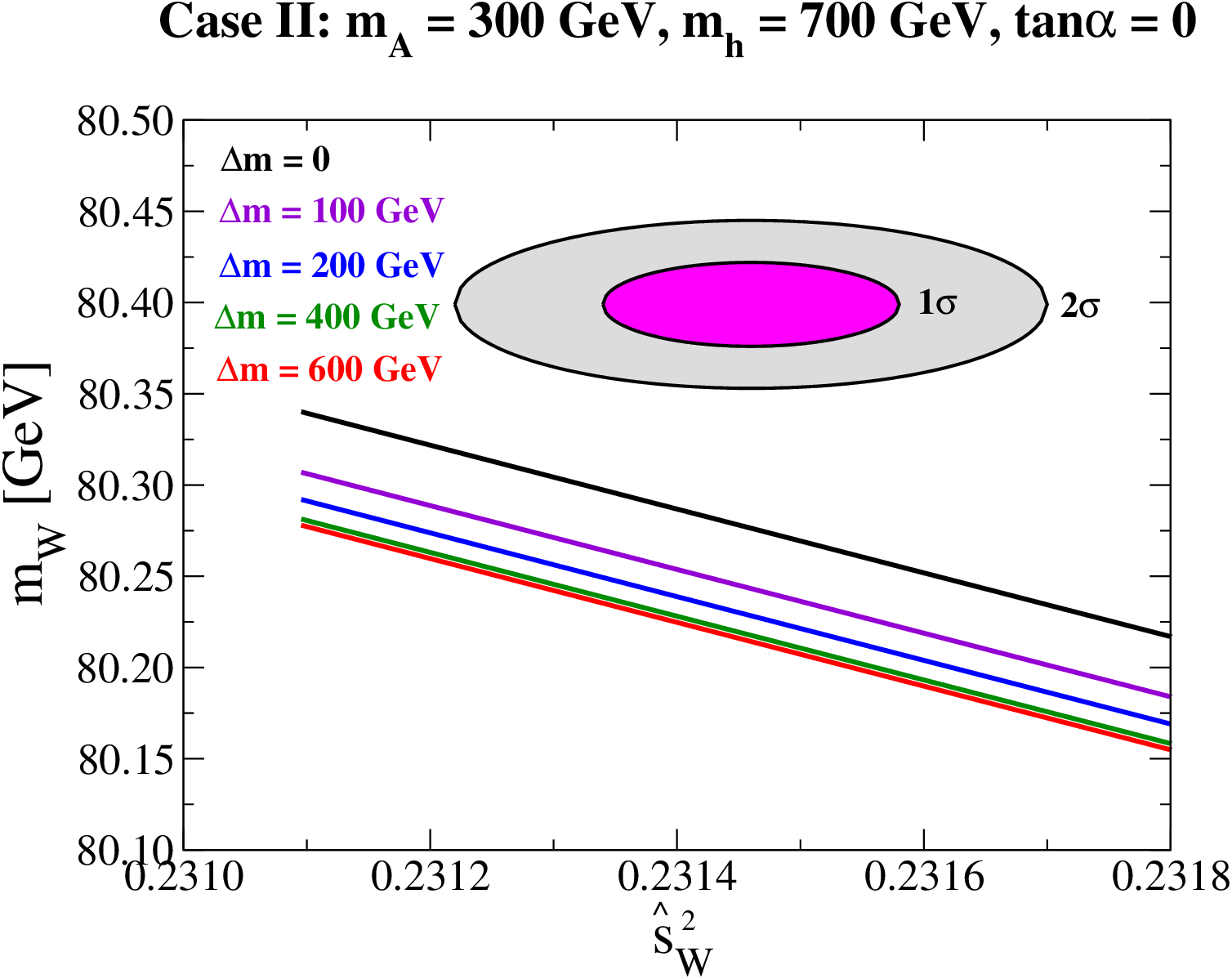}
\end{center}
\caption{The one-loop corrected values of $m_W$ as a function of $\hat{s}_W^2$ in Case II ($m_{H^{++}}>m_{H^+}>m_{\phi^0}$). 
We take $m_t=173$ GeV and $\tan\alpha=0$ in all the figures. 
The pink (gray) shaded region represents the 1$\sigma$ (2$\sigma$) error for the experimental data of 
$m_W^{\text{exp}}$ (=80.399 $\pm$ 0.23 GeV~\cite{pdg}) and $\hat{s}_W^{2\text{ exp}}$ (=0.23146 $\pm$ 0.00012 GeV~\cite{pdg}). 
In the left (right) two figures, we take $m_h=$ 125 GeV (700 GeV). 
The mass of the lightest triplet-like Higgs boson is taken to be 150 GeV and 300 GeV. }
\label{mw4}
\end{figure}

In Fig.~\ref{mw4}, 
we show the renormalized values for $m_W$ for each value of $\Delta m$ as a function of the input parameter $\hat{s}_W^2$ in Case II.  
The values of $(m_A,m_h)$ are taken to be (150 GeV,125 GeV), (300 GeV,125 GeV), (150 GeV,700 GeV) and (300 GeV,700 GeV) 
for the figures located at the upper left, the upper right, the lower left and the lower right, respectively. 
In all figures, the mixing angle $\alpha$ is set to be zero. 
Regions indicated by the data within the $1\sigma$ error and the $2\sigma$ error are 
also shown in each figure for the fixed value of $m_t$ (=173 GeV). 
For $m_{A}=150$ GeV and $m_h=125$ GeV (upper left), 
the predicted values for $m_W$ with non-zero $\Delta m$ ($>0$) are 
smaller than that with $\Delta m=0$. 
The situation is unchanged for the other choices of 
($m_A$,$m_h$)$=$(300 GeV,125 GeV), (150 GeV,700 GeV) and (300 GeV,700 GeV). 
Therefore, the hierarchical scenario with non-zero $\Delta m$ is 
highly constrained by the combined data of $m_W$ and $\hat{s}_W^2$. 
The edge of each curve at $\hat{s}_W^2\simeq 0.2311$ corresponds to $v_\Delta=0$. 
We note that the result is almost unchanged even if we vary $\tan\alpha$ in the region of $0<\tan\alpha<1$ as long as 
$A$ is not too heavy. 

We here give a comment on a Higgs sector with 
a real triplet field ($Y=0$), which is deduced from little Higgs models~\cite{little_Higgs}. 
In this Higgs sector, a CP-even scalar boson and pair of singly-charged scalar 
bosons appear in addition to the SM-like Higgs boson. 
The radiative corrections to the W boson mass have been studied in Refs.~\cite{blank_hollik,real-triplet,Chen:2005jx}. 
Similar effects of the loop corrections to the W boson mass as in Fig.~(\ref{mw3}) and Fig.~(\ref{mw4}) 
can be seen in this Higgs sector, namely 
when the mass of the SM-like Higgs boson is getting larger then the W boson mass is also getting 
larger. 
In addition, when the singly-charged scalar boson is lighter (heavier) than the additional CP-even scalar boson,  
which is a scenario such like Case I (Case II) in the HTM, 
the loop effects of these scalar bosons increase (decrease) the W boson mass~\cite{blank_hollik, Chen:2005jx}.

In Fig.~\ref{rho}, the deviation in the one-loop corrected rho parameter in the HTM
from that of the SM one-loop prediction ($\Delta \rho\equiv \rho-\rho_{\text{SM}}(m_h^{\rm ref})$)
is shown as a function of $v_\Delta$, 
where $v_\Delta$ is defined in Eq.~(\ref{data}). 
In order to describe the allowed region of $\Delta \rho$, 
we employ the data for the electroweak $T$ parameter~\cite{Peskin-Takeuchi} of 
$T=0.07\pm 0.08$~\cite{pdg}, in which $T=0$ is chosen for the reference value of the SM Higgs boson mass $m_h^{\rm ref}$ 
to be 117 GeV and $m_t=173$ GeV. 
We then obtain $\Delta\rho^{\text{exp}}=0.000632\pm 0.000621$, where $m_h^{\rm ref}=125$ GeV is chosen by taking into account 
the recent direct search results at the LHC~\cite{LHC-Higgs}.  
In the left figure, the results in Case I are shown, while in the right figure 
those in Case II are shown for several values of $\Delta m$.  
The mass of the SM-like Higgs boson is taken to be $m_h=125$ GeV, and the mixing angle $\tan\alpha$ is set to be zero. 
In Case I (left figure), the predicted values of $\Delta \rho$ for $\Delta m=0$ are 
outside the region within the $2\sigma$ error under the data 
$\Delta\rho^{\text{exp}}$ and $\hat{s}_W^2$ given in Eq.~(\ref{data}) with $m_t=173$ GeV. 
But the effect of non-zero $|\Delta m|$ makes $\Delta\rho$ larger. 
The allowed value for $v_\Delta$ within the $2\sigma$ error is 
about 3.5 GeV $\lesssim v_\Delta\lesssim 8$ GeV for about 100 GeV $\lesssim |\Delta m|\lesssim 440$ GeV.  
Notice that, as shown in Fig.~\ref{mw3}, 
the favored value of $|\Delta m|$ from the data of $m_W$ and $\hat{s}_W^2$ is about 
$200$ GeV $\lesssim|\Delta m|\lesssim 600$ GeV in Case I.  
Therefore,  
we may conclude that the combined data indicate the favored value of $v_\Delta$ to be 3.5-8 GeV in Case I 
with $m_{H^{++}}=150$ GeV. 
Next, the result in Case II is shown in the right figure, where 
the effect of $\Delta m$ ($>0$) gives the negative contribution to $\Delta \rho$. 
However, it can be seen that 
Case II is already highly constrained by the data of $m_W$ and $\hat{s}_W^2$ 
with $m_A=150$ GeV. 

\begin{figure}[t]
\begin{center}
\includegraphics[width=85mm]{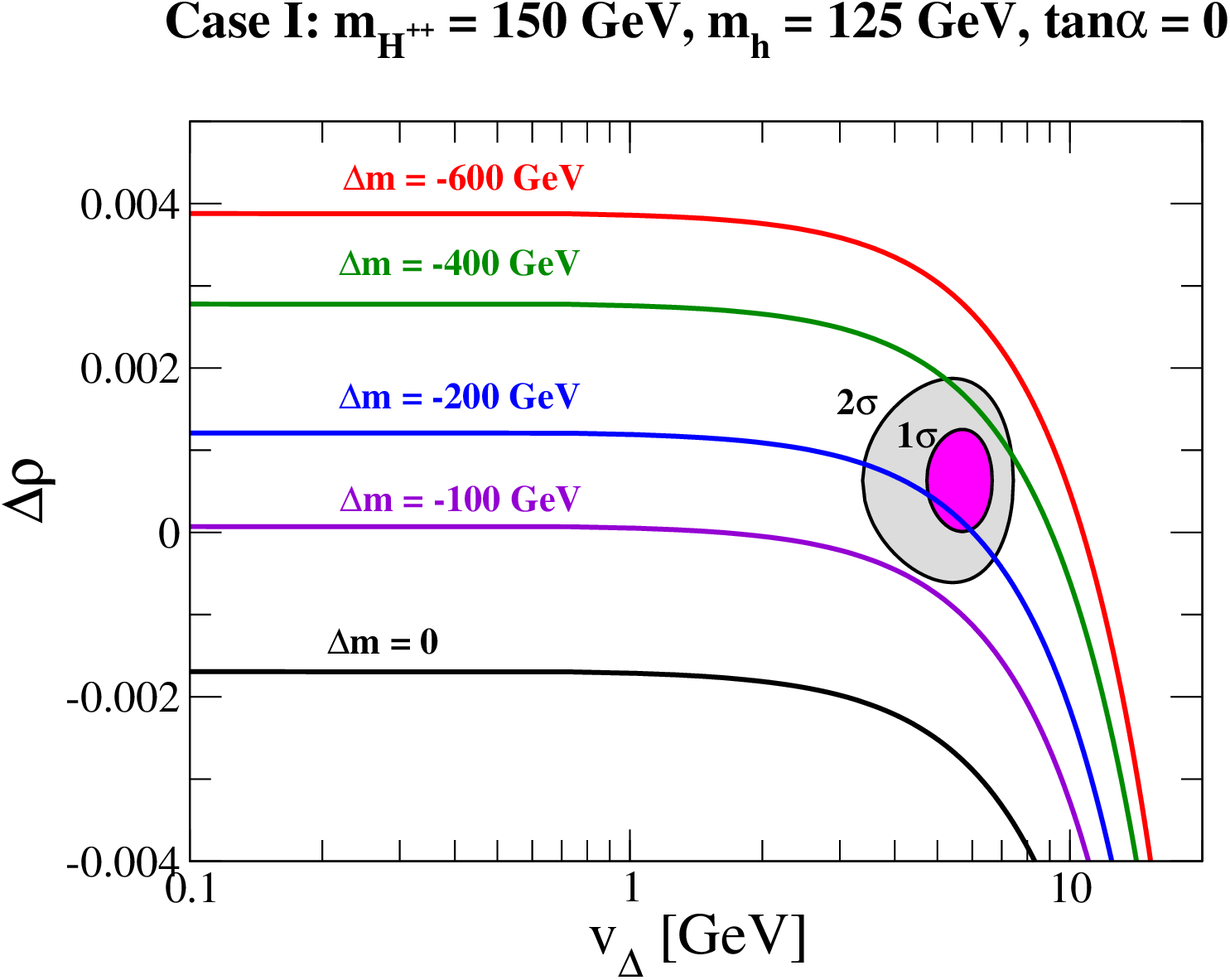}\hspace{3mm}
\includegraphics[width=85mm]{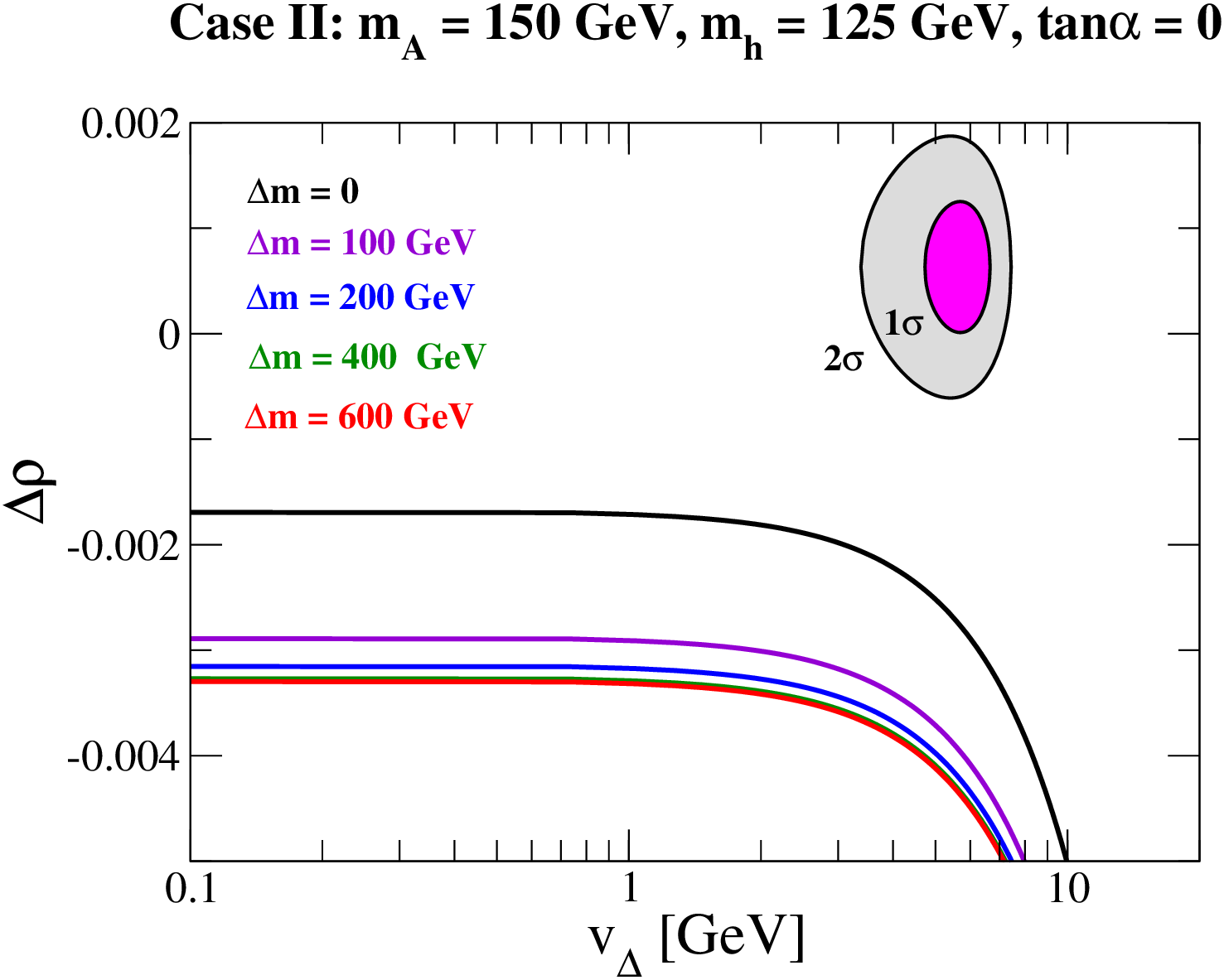}
\end{center}
\caption{The deviation of the one-loop corrected values of the rho parameter from 
those of the SM one-loop prediction 
as a function of $v_\Delta$. 
We take $m_h=125$ GeV, $m_t=173$ GeV and $\tan\alpha=0$ in the both figures. 
The pink (gray) shaded region represents the 1$\sigma$ (2$\sigma$) error for the experimental data of 
$\Delta \rho^{\text{exp}}$ (=0.000632$\pm$0.000621) which is derived from the data of the $T$ parameter (=0.07$\pm$0.08~\cite{pdg}). 
In the left figure, the mass hierarchy of the triplet-like Higgs bosons is taken to be Case I 
($m_{\phi^0} > m_{H^{+}}>m_{H^{++}}$), and $m_{\phi^0}$ is fixed to be 150 GeV. 
In the right figure, the mass hierarchy of the triplet-like Higgs bosons is taken to be Case II 
($m_{H^{++}}>m_{H^+}>m_{\phi^0}$), and $m_{H^{++}}$ is fixed to be 150 GeV.}
\label{rho}
\vspace{8mm}
\begin{center}
\includegraphics[width=90mm]{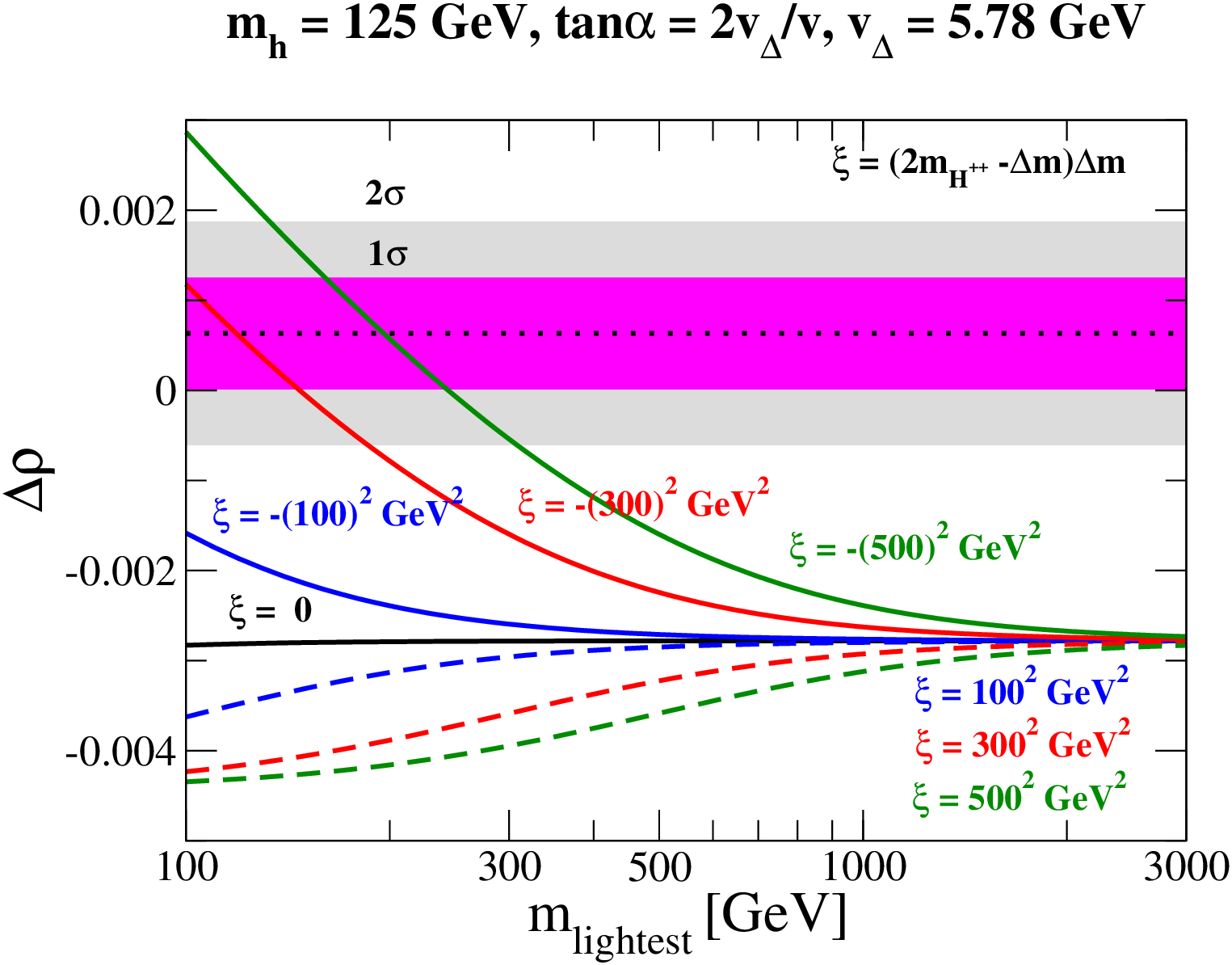}
\end{center}
\caption{The deviation of the one-loop corrected values of the rho parameter from 
those of the SM one-loop prediction 
as a function of the mass of the lightest triplet-like Higgs boson $m_{\text{lightest}}$ 
for each fixed value of $\xi$ ($\equiv m_{H^{++}}^2-m_{H^+}^2$). 
We take $m_t=173$ GeV, $m_h=125$ GeV, $v_\Delta=5.78$ GeV and $\tan\alpha=2v_\Delta/v$. 
The pink (gray) shaded region represents the 1$\sigma$ (2$\sigma$) error for the experimental data of 
$\Delta \rho^{\text{exp}}$ (=0.000632$\pm$0.000621) which is derived from the data of the 
$T$ parameter (=0.07$\pm$0.08~\cite{pdg}). 
}
\label{rho_as}
\end{figure}
We here give a comment on the decoupling property of the heavy triplet-like Higgs bosons in the electroweak observables. 
In Fig.~\ref{rho_as}, we show $\Delta\rho$ as a function of the lightest of all the triplet-like Higgs bosons for each value $\xi$ 
($\equiv m_{H^{++}}^2-m_{H^+}^2$). 
We again take $m_h=125$ GeV and $m_t=173$ GeV. 
The VEV $v_\Delta$ is fixed to be the central value indicated by the data ($v_\Delta=5.78$ GeV). 
In this figure, $\tan\alpha$ is chosen to be $2v_\Delta/v$, 
which is the asymptotic value in the limit of $m_{\text{lightest}}\to\infty$. 
It can be seen that the one-loop contribution of these particles decouples in the large mass limit. 
The asymptotic value in this limit is determined in the renormalization scheme 
with the four input parameters $\alpha_{\text{em}}$, $G_F$, $m_Z$ and $\hat{s}_W^2$ without 
the tree level relation of $m_W^2=\hat{c}_W^2 m_Z^2$. 
Therefore, it is not surprising that the asymptotic value in the HTM does not coincide with the SM value 
($\Delta\rho=0$ in this figure) with the three input parameters with $m_W^2=\hat{c}_W^2 m_Z^2$. 
In the SM and all the models with $\rho=1$ at the tree level, 
$\delta \rho$ ($\equiv \rho-1$) measures the violation of the custodial $SU(2)$ symmetry~\cite{custodial1,custodial2}. 
Such effects 
appear as the quadratic power-like contributions of the mass difference between particles 
in the $SU(2)$ multiplet; i.e., 
$\delta\rho\simeq$ $(m_u-m_d)^2/v^2$ for $m_u\simeq m_d$ via the chiral fermion loop diagram~\cite{fermion_loop}, 
and $\delta\rho\simeq$ $(m_{H^+}-m_A)^2/v^2$ for $m_{H^+}\simeq m_A$ via the additional scalar boson loop diagram~\cite{
Toussaint,Bertolini,Peskin_Wells,Osland,Taniguchi} 
in the general two Higgs doublet model.
On the other hand, in the models with $\rho\neq 1$ at the tree level like the HTM, 
such quadratic power-like mass contributions are absorbed by 
renormalization of the new independent input parameter $\delta \hat{s}_W^2$.  
Consequently, only a logarithmic dependence on the masses of the particles in the 
loop diagram remain. 
In other words, in these models the rho parameter is no more the parameter which measures the violation 
of the $SU(2)$ custodial symmetry in the sector of particles in the loop. 
This has already been known in the calculation of the model with the $Y=0$ triplet field~\cite{blank_hollik,Chen:2005jx}\footnote{
In Ref.~\cite{Chen:2005jx}, it is claimed that quadratic mass contributions appear. 
However, their approximate formulae for $\Delta r$ seem to be inconsistent. 
Quite recently this was also pointed out in Ref.~\cite{Yale}.}.

\section{Implication to the Higgs searches at the LHC}

In this section, we discuss several phenomenological consequences of constraints from 
the electroweak precision data in the HTM which is discussed in the previous sections. 
Recent results for the Higgs boson searches at the LHC indicate that 
the Higgs boson mass is between 115 GeV and 127 GeV at the 95\% confidence level 
assuming that the Higgs boson is of SM-like~\cite{LHC-Higgs}. 
The Higgs boson $\phi$ is tested mainly via the decay processes such as 
$\phi\to \gamma\gamma$, $\phi\to ZZ^\ast$, $\phi\to WW^\ast$ and $\phi\to \tau^+\tau^-$. 
First of all, we discuss the radiative effect of triplet-like Higgs bosons on 
the decay rate of $h\to \gamma\gamma$ in the HTM under the constraint from the electroweak precision data. 
The $h\gamma\gamma$ vertex is generated at the one-loop level,  
so that the significant one-loop contributions of additional charged particles can appear. 
In the HTM, there are doubly- and singly-charged Higgs bosons which would give substantial 
one-loop contributions to the decay rate of $h\to \gamma\gamma$. 
Recently, Arhrib et.~al have discussed details of this decay process in the HTM under the constraint from 
perturbative unitarity and vacuum stability~\cite{Arhrib}. 
We here analyze the decay rate taking into account our new results of the radiative corrections to the 
electroweak parameters.

\begin{figure}[t]
\begin{center}
\includegraphics[width=85mm]{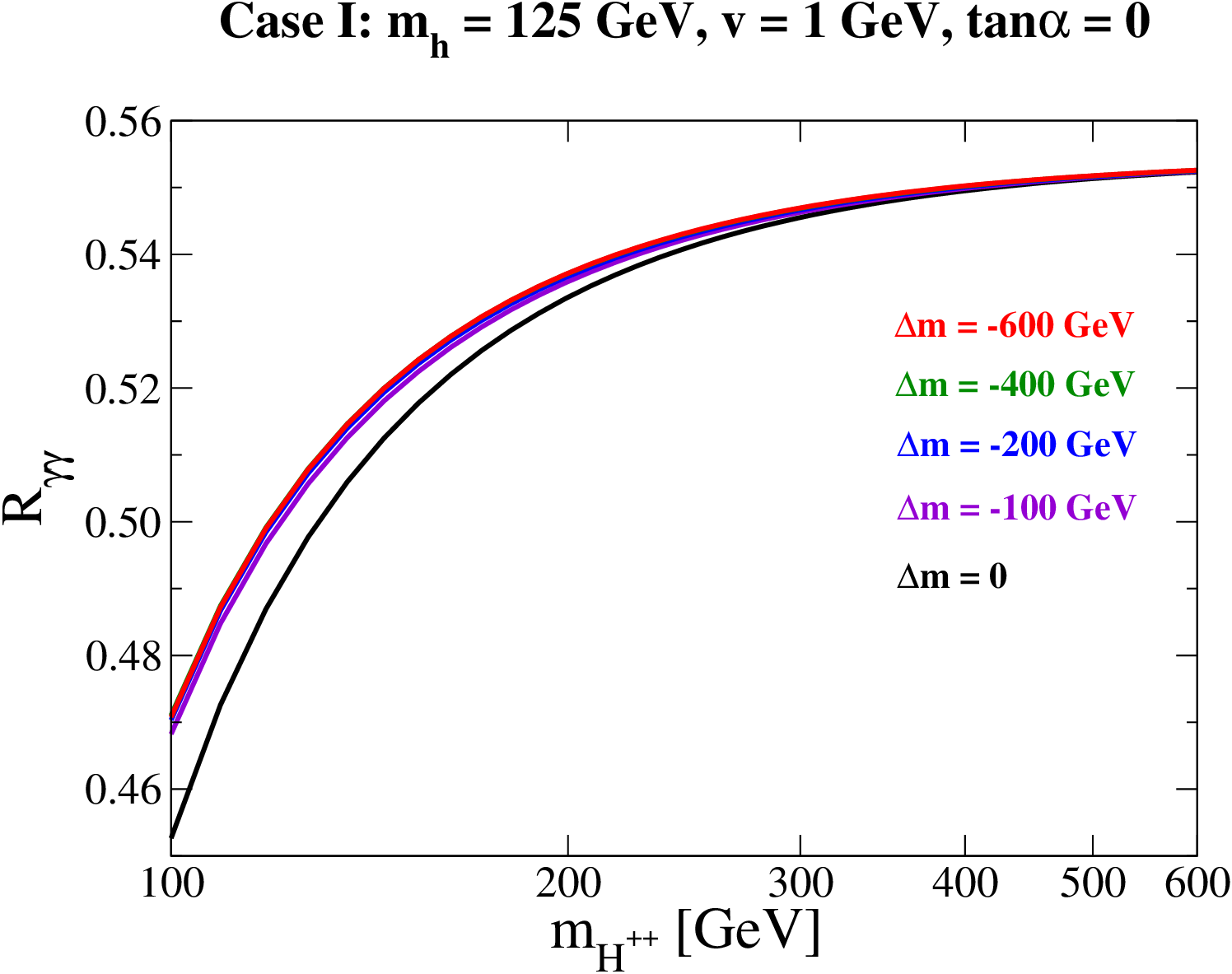}\hspace{3mm}
\includegraphics[width=85mm]{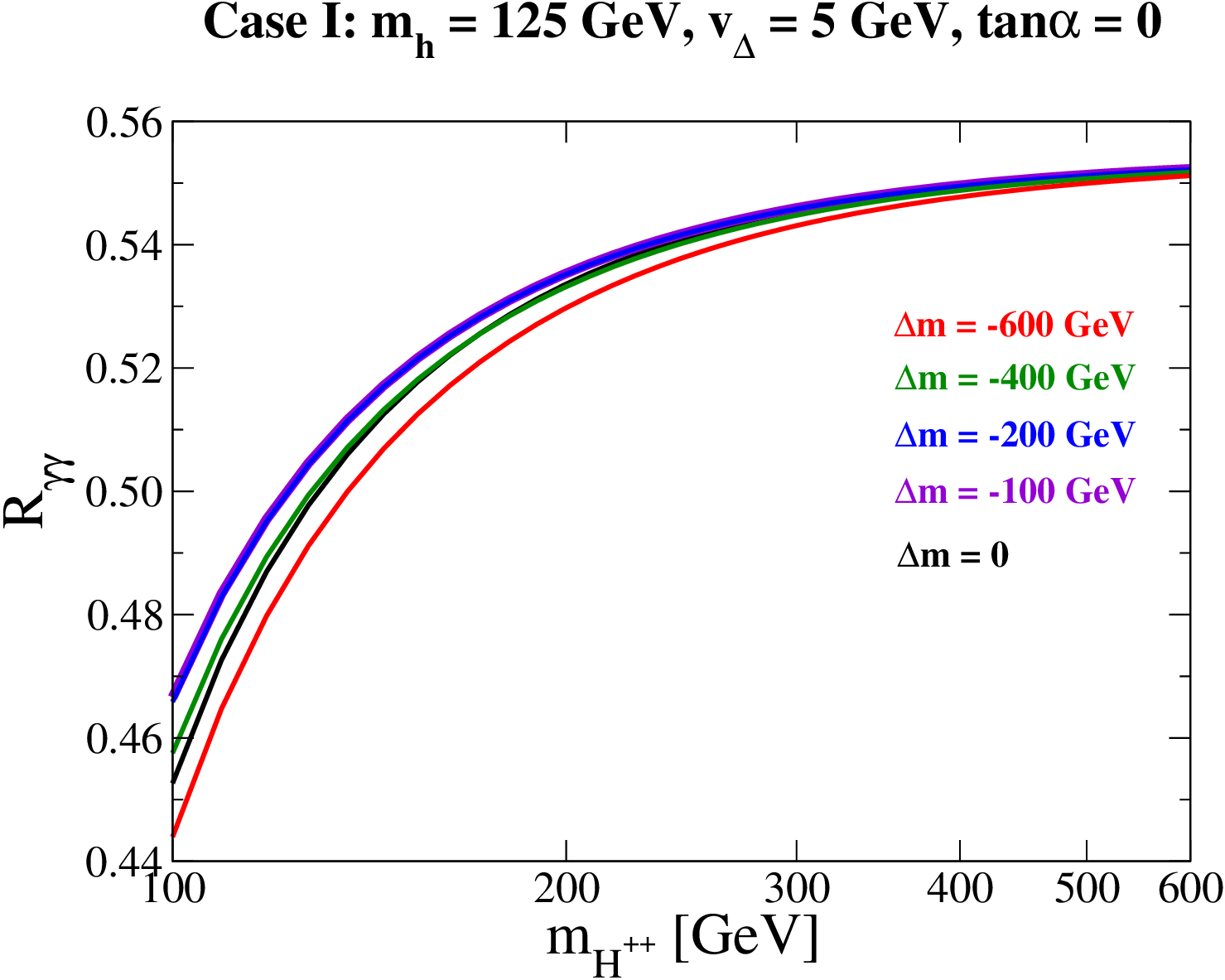}
\end{center}
\caption{The ratio of the decay rate for $h\to \gamma\gamma$ in the HTM to that in the SM as a function of $m_{H^{++}}$ 
for each fixed value of $\Delta m$ ($<0$) in Case I $(m_{\phi^0}>m_{H^+}>m_{H^{++}})$. 
In the both figures, we take $m_t=173$ GeV, $m_h=125$ GeV and $\tan\alpha=0$. 
In the left (right) figure, we take $v_\Delta=1$ GeV (5 GeV).}
\label{hgg1}
\end{figure}

\begin{figure}[t]
\begin{center}
\includegraphics[width=85mm]{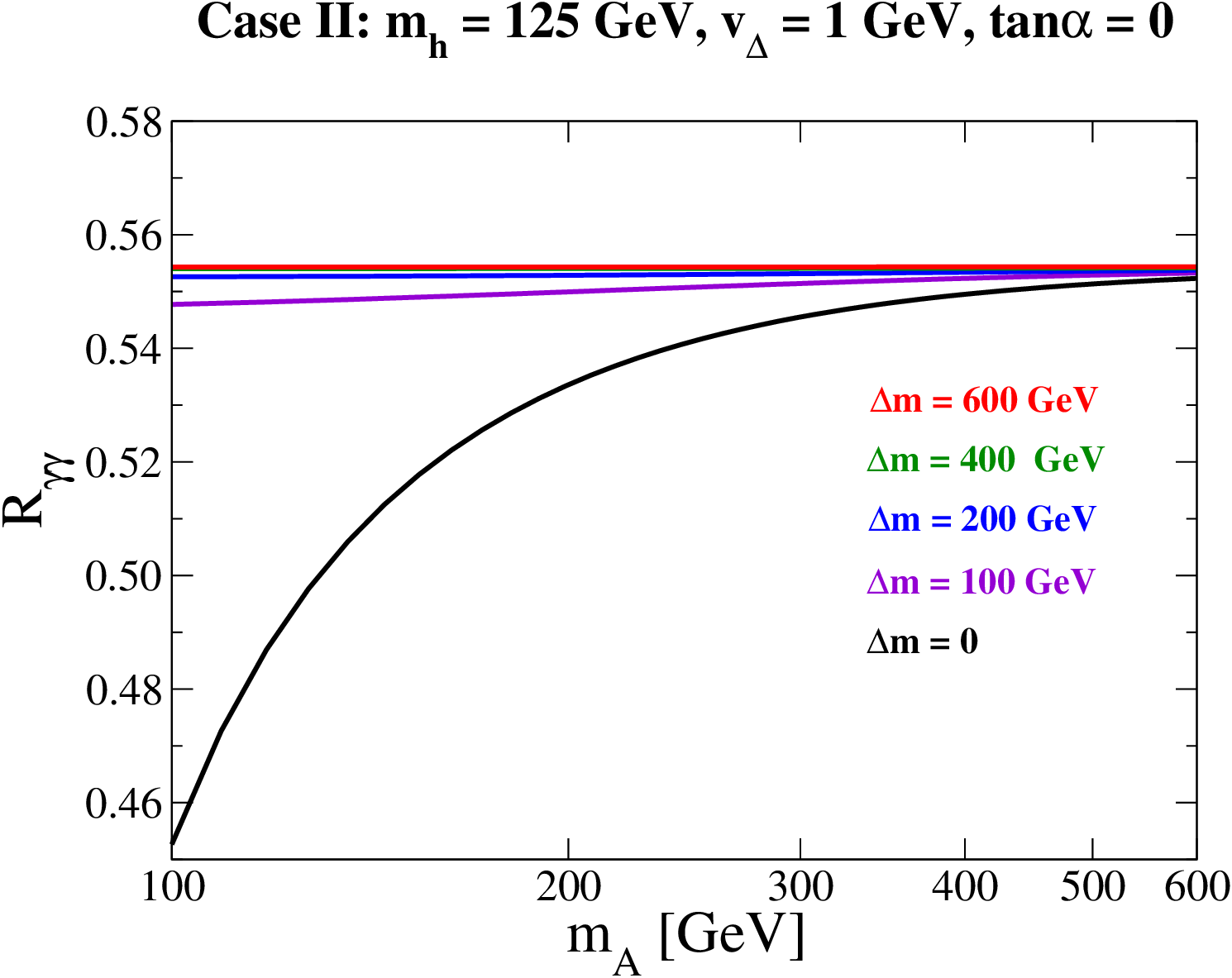}\hspace{3mm}
\includegraphics[width=85mm]{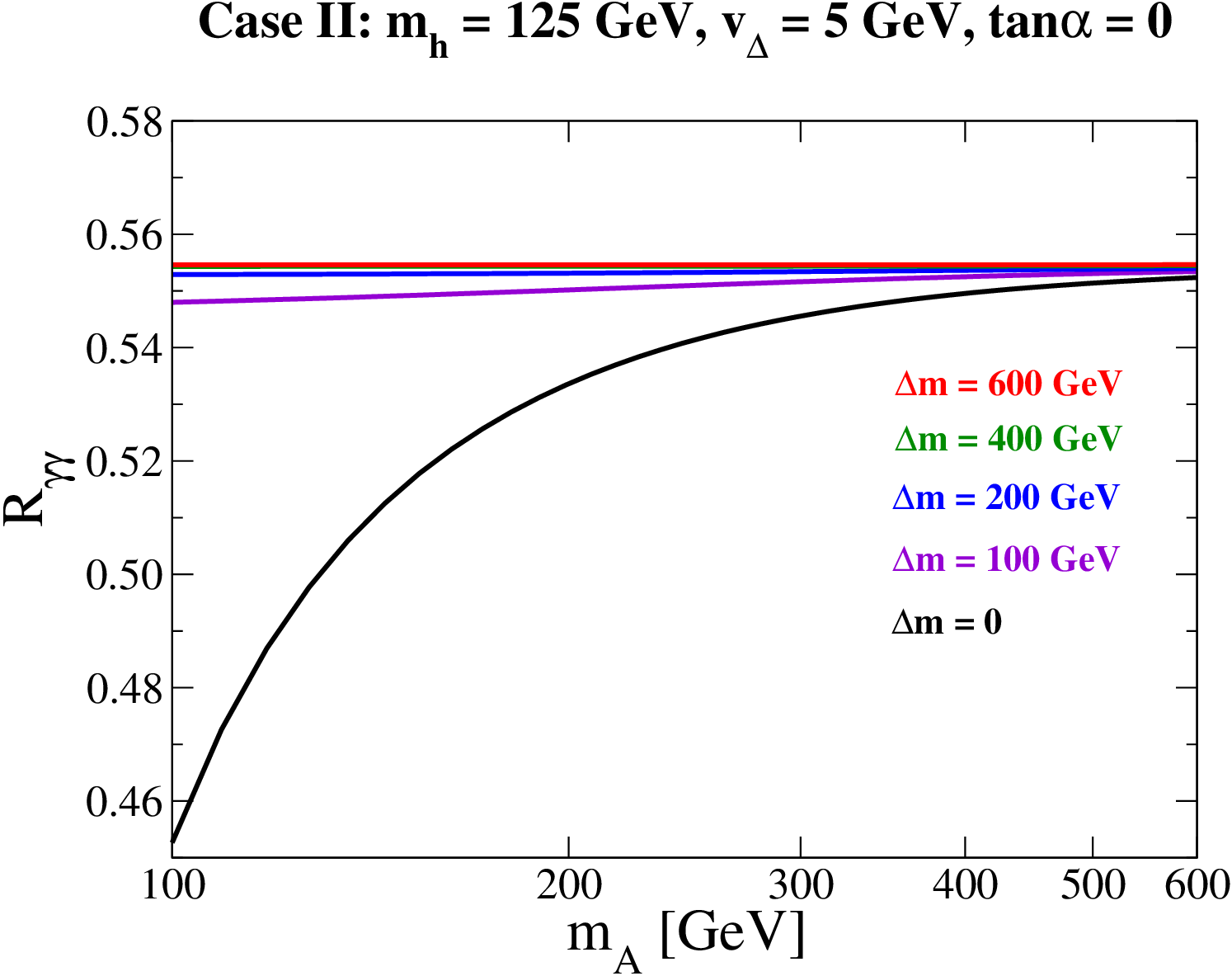}
\end{center}
\caption{The ratio of the decay rate for $h\to \gamma\gamma$ in the HTM to that in the SM as a function of $m_A$ 
for each fixed value of $\Delta m$ ($>0$) in Case II $(m_{H^{++}}>m_{H^+}>m_{\phi^0})$. 
In the both figures, we take $m_t=173$ GeV, $m_h=125$ GeV and $\tan\alpha=0$. 
In the left (right) figure, we take $v_\Delta=1$ GeV (5 GeV).}
\label{hgg2}
\end{figure}

The decay rate of $\phi\to \gamma\gamma$ is calculated at the one-loop level by~\cite{hgg}
\begin{align}
\Gamma(\phi\to \gamma\gamma)=\frac{G_F\alpha_{\text{em}}^2m_\phi^3}{128\sqrt{2}\pi^3}
\Bigg|&-2\sum_fN_f^cQ_f^2\tau_f[1+(1-\tau_f)f(\tau_f)]+2+3\tau_W+3\tau_W(2-\tau_W)f(\tau_W)\notag\\
&+Q_{H^{++}}^2\frac{2vc_{hH^{++}H^{--}}}{m_\phi^2}[1-\tau_{H^{++}}f(\tau_{H^{++}})]
+Q_{H^{+}}^2\frac{2vc_{hH^{+}H^{-}}}{m_\phi^2}[1-\tau_{H^{+}}f(\tau_{H^{+}})]\Bigg|^2, \label{hgg}
\end{align}
where the function $f(x)$ is given by 
\begin{align}
f(x)=\left\{
\begin{array}{c}
[\arcsin(1/\sqrt{x})]^2, \quad \text{if }x\geq 1,\\
-\frac{1}{4}[\ln \frac{1+\sqrt{1-x}}{1-\sqrt{1-x}}-i\pi]^2, \quad \text{if }x< 1
\end{array}\right.. 
\end{align}
In Eq.~({\ref{hgg}}),  $Q_\varphi$ is the electric charge of the field $\varphi$, 
$N_f^c$ is the color factor and $\tau_\varphi= 4m_\varphi^2/m_\phi^2$.  
In the HTM, the coupling constants $c_{h H^+H^-}$ and $c_{h H^{++}H^{--}}$ are given by 
\begin{align}
c_{h H^+H^-}&=\frac{1}{v_\Delta}\left[
m_{H^+}^2\left(\sqrt{2}s_{\beta_\pm}c_{\beta_\pm}c_\alpha+2s_{\beta_\pm}^2s_\alpha\right)
-m_A^2s_\alpha\left(c_{\beta_0}^2+\frac{s_{\beta_0}^2}{2}\right)
+m_h^2\left(\frac{s_{\beta_\pm}^3c_\alpha}{\sqrt{2}c_{\beta_\pm}}+c_{\beta_\pm}^2s_\alpha\right)\right],\\
c_{h H^{++}H^{--}}&=\frac{1}{v_\Delta}\left[
2m_{H^{++}}^2s_\alpha+m_h^2s_\alpha
-2m_{H^+}^2\left(2c_{\beta_\pm}^2s_\alpha-\sqrt{2}s_{\beta_\pm}c_{\beta_\pm}c_\alpha\right)
-m_A^2\left(s_{\beta_0}c_{\beta_0}c_\alpha-c_{\beta_0}^2s_\alpha\right)
\right].
\end{align}

In Fig.~\ref{hgg1}, the ratio of the decay rates 
$R_{\gamma\gamma}\equiv \Gamma(h\to \gamma\gamma)_{\text{HTM}}/\Gamma(\phi_{\text{SM}}\to \gamma\gamma)_{\text{SM}}$ 
is shown as a function of $m_{H^{++}}$ for each value of $\Delta m$ at $m_h(=m_{\phi_{\text{SM}}})=125$ GeV 
and $\tan\alpha=0$ in Case I ($m_{\phi^0}>m_{H^+}>m_{H^{++}}$). 
For the left figure, $v_\Delta$ is taken to be 1 GeV, while 
it is taken to be 5 GeV for the right figure. 
In the both figures, $R_{\gamma\gamma}<1$ because 
the one-loop contributions of the singly-charged Higgs boson and the doubly-charged Higgs boson 
to $\Gamma(\phi\to \gamma\gamma)$ have the same sign which is destructive to the contribution of the SM loop diagrams. 
The magnitude of the deviation from the SM can be significant, which amounts to larger than 40\%. 
For $v_\Delta=1$ GeV the deviation is smaller when larger $\Delta m$ is taken.   
The deviation becomes smaller and insensitive to $\Delta m$ in the large mass region for $H^{\pm\pm}$. 

\begin{figure}[t]
\begin{center}
\includegraphics[width=85mm]{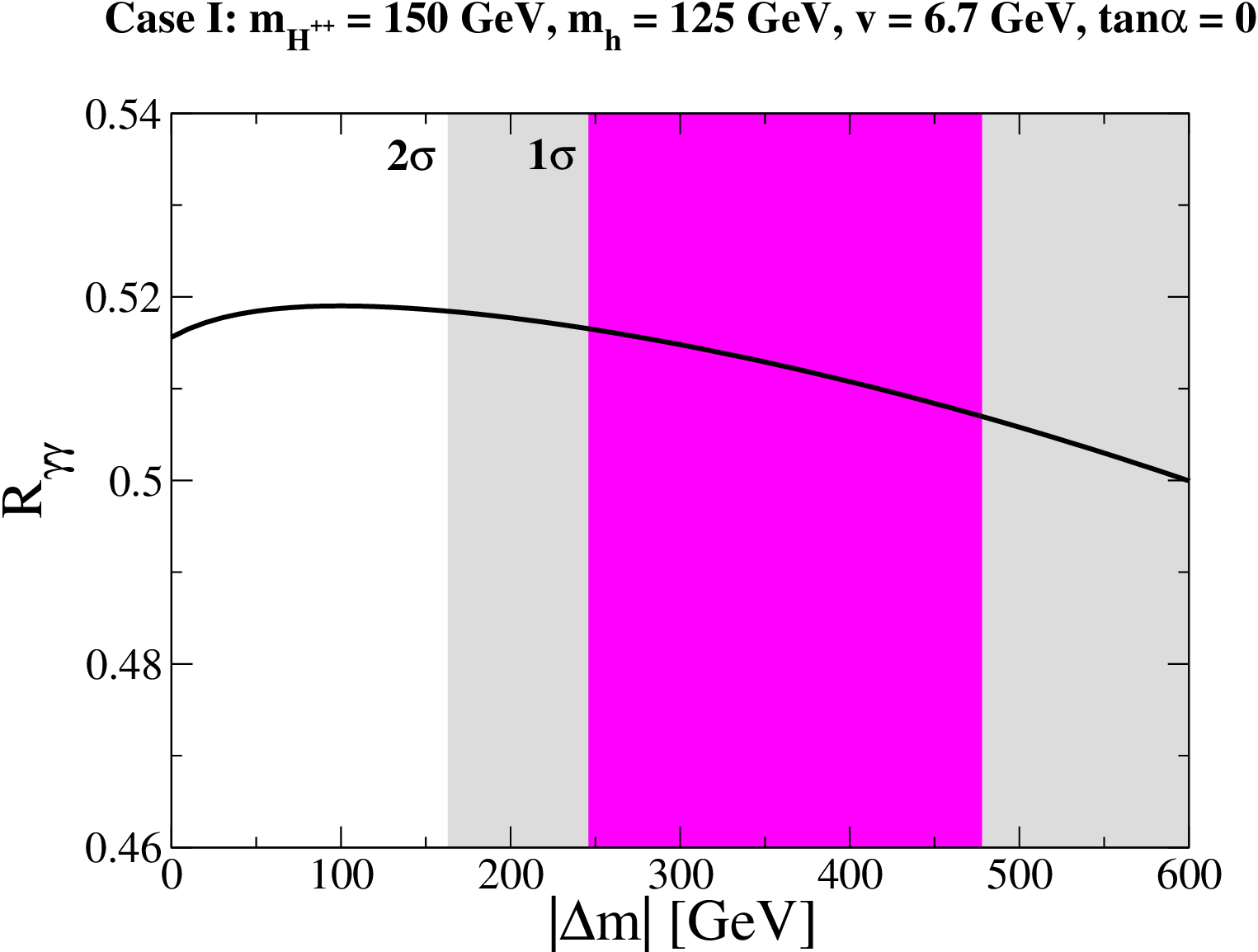}\hspace{3mm}
\includegraphics[width=85mm]{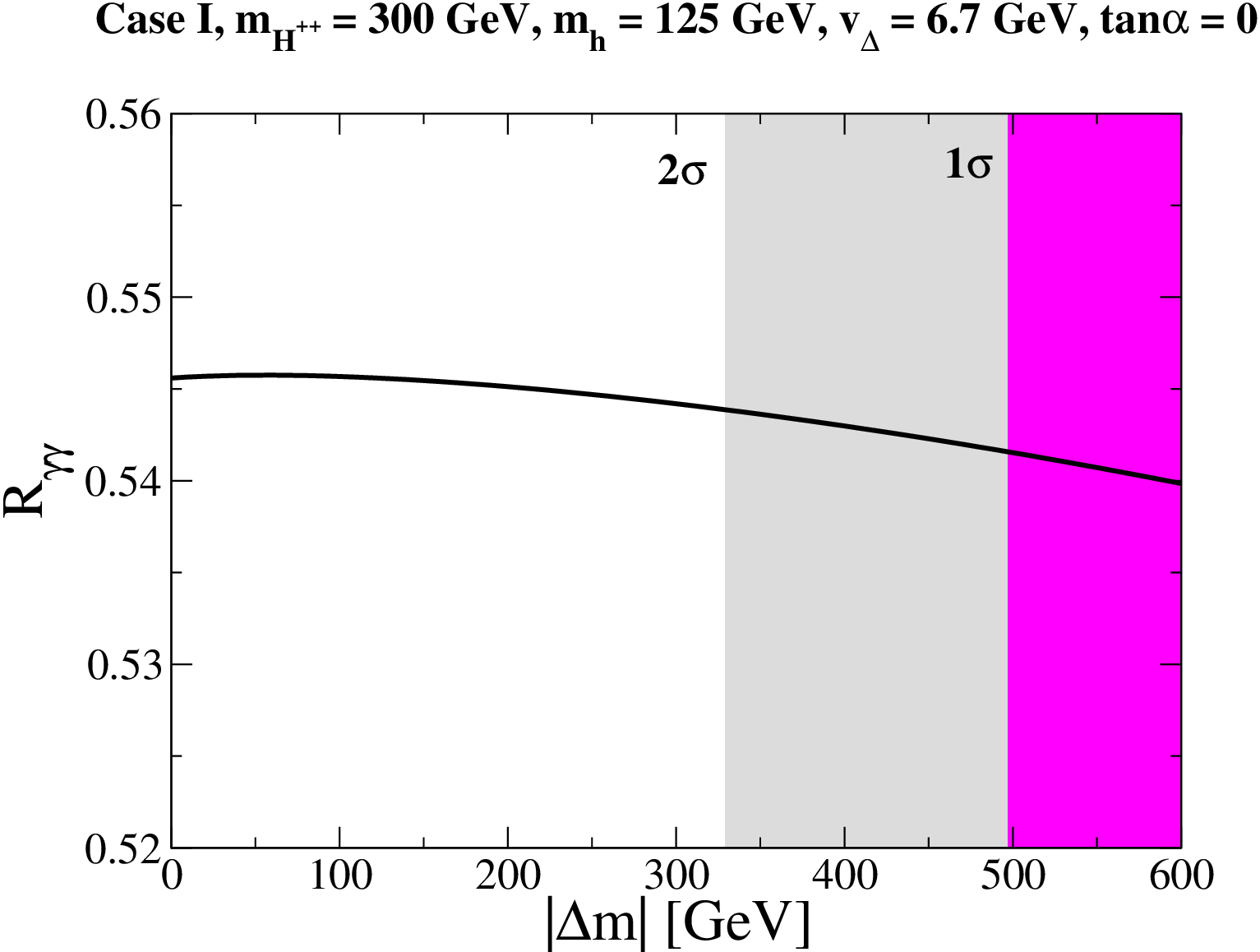}\hspace{3mm}
\end{center}
\caption{The ratio of the decay rate for $h\to \gamma\gamma$ in the HTM to that in the SM as a function of 
the absolute value of $\Delta m$ in Case I $(m_{\phi^0}>m_{H^+}>m_{H^{++}}$). 
We take $m_t=173$ GeV, $m_h=125$ GeV, $v_\Delta=6.7$ GeV and $\tan\alpha=0$ in all the figures. 
In the left (right) figure, we take $m_{H^{++}}=150$ GeV (300 GeV). 
The pink (gray) shaded region represents the 1$\sigma$ (2$\sigma$) allowed region of $\Delta m$ under the constraint from 
the data for $m_W^{\text{exp}}$ and $\hat{s}_W^{2\text{ exp}}$.
}
\label{hgg3}
\end{figure}

One might think that the deviation would approach to zero in the large mass limit for $H^{\pm\pm}$. 
This can be true in a generic case. 
However, such decoupling is not applied to the present case. 
Since the coupling constants $c_{hH^{+}H^{-}}$ and $c_{hH^{++}H^{--}}$ are both proportional to 
the mass squired of triplet-like Higgs bosons, 
the large mass limit with a fixed value of $\Delta m$ with $\alpha=0$ can only be realized by taking 
these coupling constants to be infinity. 
It is known that in such a case, Appelquist's decoupling theorem~\cite{Appelquist} does not hold, 
and the one-loop contributions of $H^\pm$ and $H^{\pm\pm}$ remain 
in the large mass limit as non-decoupling effects. 
We note that,   
in this case with $\alpha=0$, we have the relation $m_A^2\simeq M_\Delta^2=(\lambda_4+\lambda_5)v_\Phi^2/2$ from Eq.~(\ref{tan2a}), 
so that all the masses of triplet-like Higgs bosons cannot be taken to be  larger than 
TeV scales because of the perturbative unitarity constraint. 
On the contrary, 
if $\alpha=0$ is relaxed, 
$m_A$ is a free parameter, which satisfies 
$m_A^2\simeq M_\Delta^2=(\mu/v_\Delta)v_\Phi^2/\sqrt{2}$ from Eq.~(\ref{vc}),  
and it can be taken to be much larger than the electroweak scale when   
$\mu/v_\Delta\gg 1$ is assumed. 
Then, the correction due to the triplet field is suppressed by a factor of 
$v^2/m_A^2$. 
Namely, the decoupling theorem holds in this case.

In Fig.~\ref{hgg2}, $R_{\gamma\gamma}$ is shown as a function of $m_A$ 
for each value of $\Delta m$ at $m_h(=m_{\phi_{\text{SM}}})=125$ GeV 
and $\alpha=0$ 
in Case II ($m_{H^{++}}>m_{H^+}>m_{\phi^0}$). 
It is seen that as compared to Case I $R_{\gamma\gamma}$ 
is sensitive to the choice of $\Delta m$. 
Similarly to Case I, the deviation from the SM value is negative. 
However, smaller deviation is obtained for larger $\Delta m$ for the both cases with 
$v_\Delta=1$ GeV and $v_\Delta=5$ GeV in the region of relatively lower values of $m_A$.

In Fig.~\ref{hgg3}, we show the results of $R_{\gamma\gamma}$ as a function of $|\Delta m|$ 
in Case I with indicating the allowed regions of each confidence level under the 
electroweak precision data. 
The mass of $H^{\pm\pm}$ is taken to be 150 GeV (left) and 300 GeV (right). 
In all the figures, we take $m_h=125$ GeV, $\tan\alpha=0$ and $v_\Delta=6.7$ GeV. 
The magnitude of the ratio $R_{\gamma\gamma}$ strongly depends on $m_{H^{++}}$. 
Therefore, we may be able to obtain the indirect information of the mass spectrum of the 
triplet-like Higgs bosons by measuring the decay rate of $h\to \gamma\gamma$.

\begin{figure}[t]
\begin{center}
\includegraphics[width=85mm]{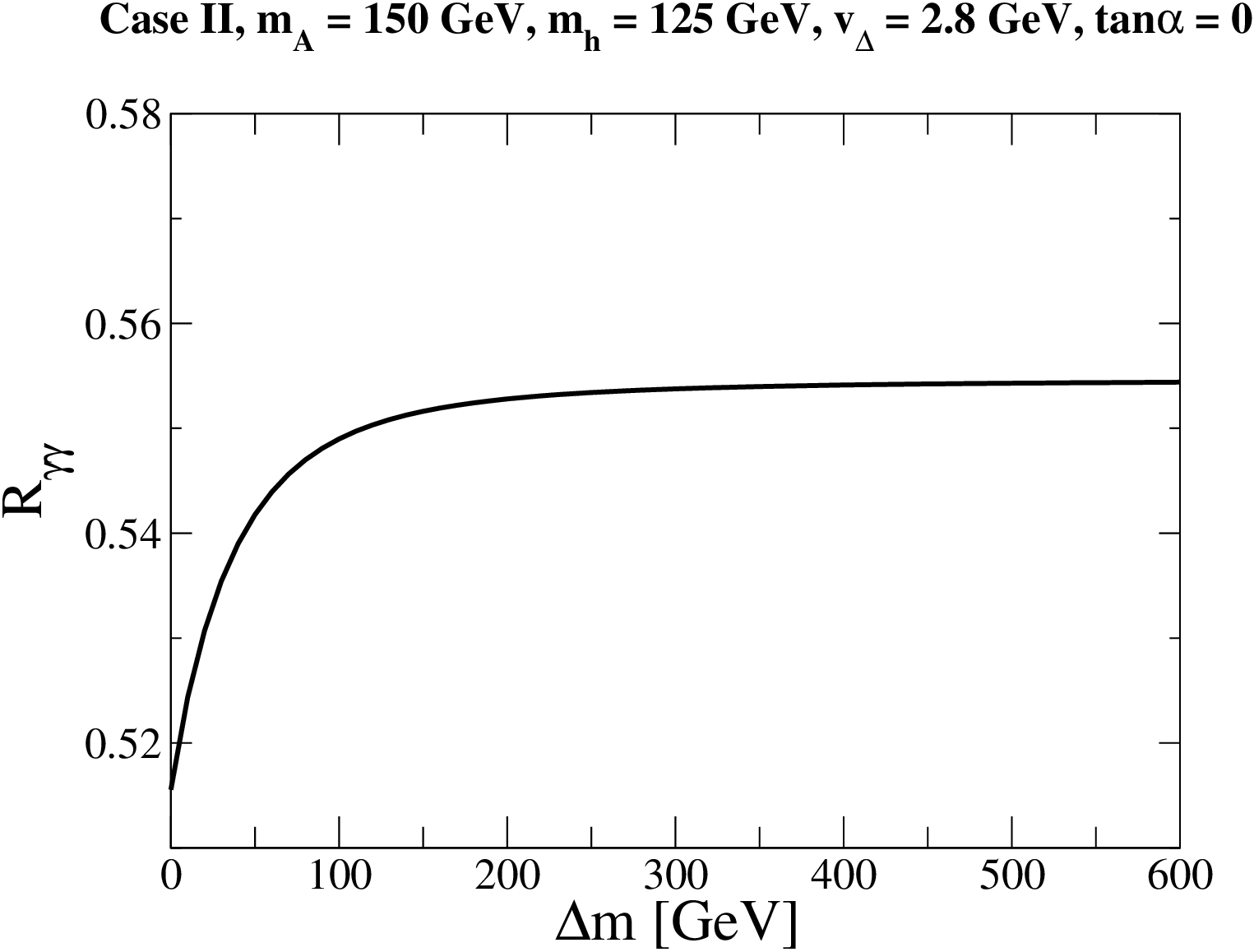}\hspace{3mm}
\includegraphics[width=85mm]{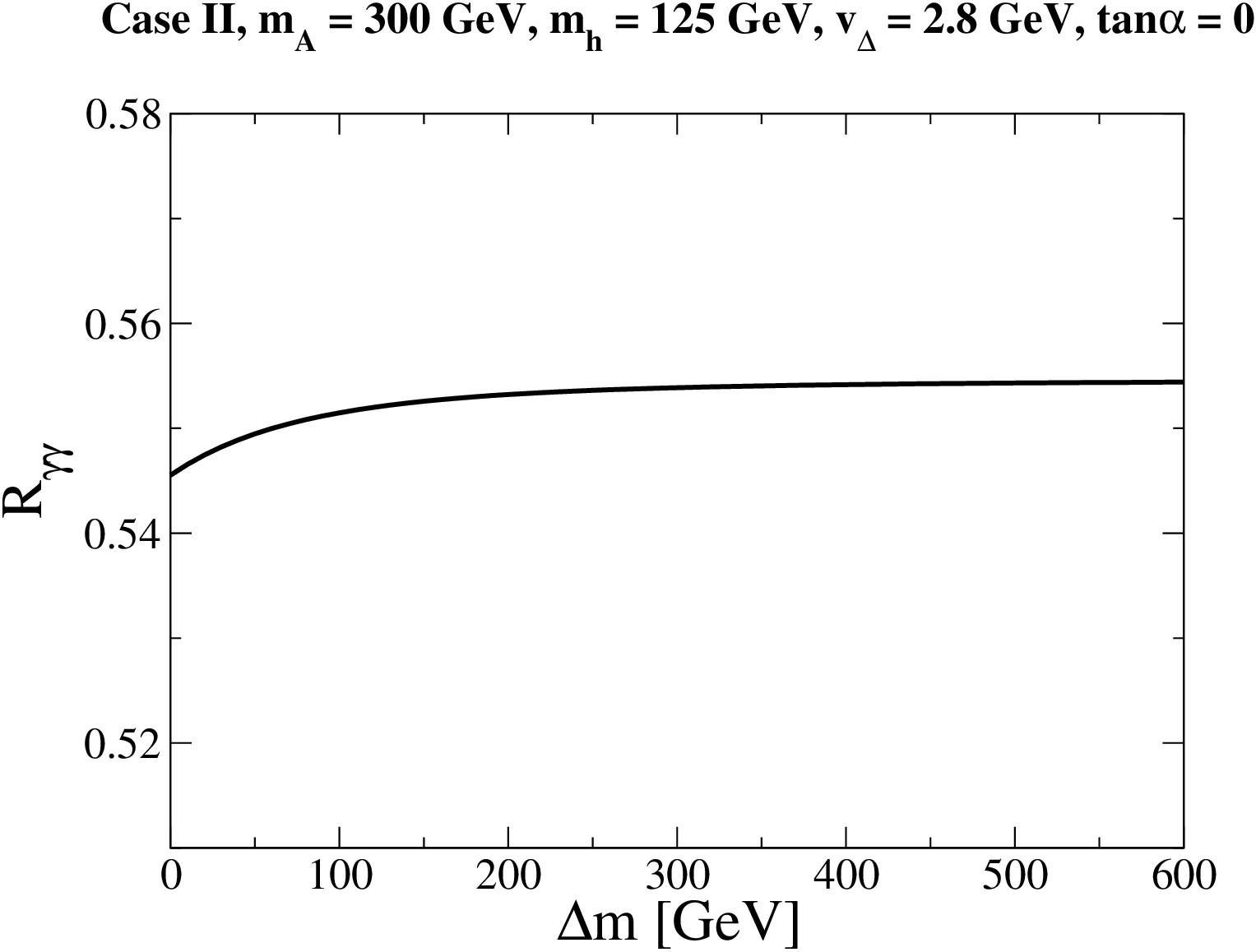}\hspace{3mm}
\end{center}
\caption{The ratio of the decay rate for $h\to \gamma\gamma$ in the HTM to that in the SM as a function of 
$\Delta m$ in Case II $(m_{H^{++}}>m_{H^+}>m_{\phi^0})$. 
We take $m_t=173$ GeV, $m_h=125$ GeV, $v_\Delta=2.8$ GeV and $\tan\alpha=0$ in all the figures. 
In the left (right) figure, we take $m_{A}=150$ GeV (300 GeV). 
There is no consistent region whth the data for $m_W^{\text{exp}}$ and $\hat{s}_W^{2\text{ exp}}$. }
\label{hgg4}
\end{figure}

Finally, in Fig.~\ref{hgg4}, $R_{\gamma\gamma}$ is shown as a function of $\Delta m$ 
in Case II with indicating the allowed regions of each confidence level under the 
electroweak precision data. 
The mass of $A$ is taken to be 150 GeV (left) and 300 GeV (right). 
In all the figures, we take $m_h=125$ GeV, $\tan\alpha=0$ and $v_\Delta=2.8$ GeV. 
As compared to the case shown in Fig.~\ref{hgg3}, 
the mass dependence on $m_A$ is small among the three values of $m_A$. 
As we already discussed, Case II is not preferred by the electroweak precision data, 
and only the region with larger deviation than $2\sigma$ appears in each figure.

In the following, we present several comments on the prospect for direct searches of 
the triplet-like Higgs bosons at the LHC. 
The most important experimental signature for the $Y=1$ HTM would be the detection of the doubly-charged Higgs bosons. 
For the degenerate scenario ($m_{H^{++}}\simeq m_{H^+}\simeq m_A\simeq m_H$) and Case I, 
$H^{\pm\pm}$ can decay into $\ell^\pm\ell^\pm$ for a relatively small value of $v_\Delta$ 
and $W^\pm W^\pm$ for larger $v_\Delta$~\cite{Gunion_Olness,Gunion_Vega_Wudka}. 
If $m_{H^{++}}=120$ GeV (300 GeV) the crossing point is at $v_\Delta\sim \mathcal{O}(10)$ MeV 
($v_\Delta\sim \mathcal{O}(0.1)$ MeV). 
Comprehensive phenomenological study has been done in Refs.~\cite{TaoHan} for the 
degenerate scenario, and in Ref.~\cite{Akeroyd-Sugiyama} for the hierarchical scenario in Case I. 
Assuming that the dilepton decay is the main mode 
the mass of $H^{\pm\pm}$ is constrained to be $m_{H^{++}}\gtrsim 300$ GeV by the direct search results at the LHC, while 
there has been no significant constraint on $m_{H^{++}}$ when the diboson decay is dominant. 
Our results on the radiative correction to the precision parameters would prefer a non-zero value for 
$\Delta m$ with $v_\Delta$ being several GeV. 
When we assume $m_h=125$ GeV and $m_{H^{++}}=150$ GeV, 
the values of $v_\Delta$ and $\Delta m$ which are consistent with the data for $m_W$ and $\hat{s}_W^2$ 
in the $2\sigma$ region are $v_\Delta=3.5$ - 8 GeV and $|\Delta m|=160$ - 440 GeV: see Fig.~\ref{mw3}. 
In such a case, main decay modes of $H^{\pm\pm}$, $H^\pm$ and $A$ (or $H$) 
are $H^{\pm\pm}\to W^\pm W^\pm$, $H^\pm \to H^{\pm\pm}W^\mp$ and $A$ (or $H$) $\to H^{\pm}W^\mp$~\cite{Akeroyd-Sugiyama, Akeroyd_Moretti, 
Akeroyd_Moretti_Sugiyama}. 
In addition to the production processes of $q\bar{q}\to H^{++}H^{--}$ and $q\bar{q}'\to H^{\pm\pm}H^\mp$, 
more number of $H^{\pm\pm}$ is produced via the cascade decays from $HA$, $HH^\pm$, $AH^\pm$ and $H^+H^-$. 
In principle, masses of these triplet-like Higgs bosons can be measured by the endpoint analysis of the transverse mass distributions 
of parent neutral or singly-charged Higgs bosons. 
The feasibility of such a processes are however unclear, and 
realistic full simulation study is clearly required. 

Next, the hierarchical scenario of Case II seems to be more constrained than the scenarios discussed above. 
Assuming $m_h$ to be 125 GeV, there is no parameter space which is consistent with the region 
within the $2\sigma$ ellipse of $m_W^{\text{exp}}$ and $\hat{s}_W^{2\text{ exp}}$. 
A larger mass difference $\Delta m$ makes the situation worse. 
Although if the Higgs boson mass $m_h$ is rather heavy the situation can be a little bit improved, 
no consistent region at the $2\sigma$ level is found for all the mass values of $m_h$ below the unitarity bound. 
Although Case II is rather constrained from the electroweak precision data 
we may consider the phenomenology for such a case with $m_{H^{++}}>m_{H^+}>m_A$ at the LHC\footnote{
The constraint from the precision data for the minimal triplet model can be relaxed
if the Higgs sector is extended with some additional scalar bosons. 
Then 
our analysis may change by the one loop effect of new particles. }. 
In this case, the main decay mode of $H^{\pm\pm}$ can be $\ell^\pm\ell^\pm$, $H^\pm W^\pm$ or $W^\pm W^\pm$ 
depending on the magnitude of $v_\Delta$. 
In Ref.~\cite{aky}, phenomenology on this case is discussed. 
In particular, the decay modes of the doubly-charged Higgs boson such as 
$H^{\pm\pm}\to H^\pm W^\pm\to AW^\pm W^\pm$ (and $H^{\pm\pm}\to H^\pm W^\pm\to HW^\pm W^\pm$) are studied at the LHC. 

Once doubly- and singly-charged Higgs bosons as well as additional neutral Higgs bosons 
are found and their masses are measured with some mass splitting at the LHC, 
we may be able to distinguish the HTM from the other models which contain doubly-charged Higgs bosons, 
such as the Zee-Babu model~\cite{Zee-Babu}, the left-right symmetric model~\cite{LR} and 
the other exotic models~\cite{GM,Gunion_Vega_Wudka,AK_GM,aky23} by using 
the characteristic mass spectrum of the HTM given in Eqs.~(\ref{mass1}) and (\ref{mass2}). 
It would be important to examine the radiative correction to these relations. 
We will discuss this point in our next paper~\cite{akky}, where 
radiative corrections to the Higgs potential in the HTM are studied in the on-shell scheme. 

Finally, as we have seen in the previous section, when the mass of the lightest 
triplet-like field is at the electroweak scale 
the hierarchical scenario (Case I with $\Delta m<0$) is favored and the preferred value of the VEV 
$v_\Delta$ is as large as several GeV by  
the combined results for the predictions of $m_W$ and $\rho$ with the precision data.  
Such a value of $v_\Delta$ might be large enough to be directly examined by measuring the Higgs boson 
vertices which are proportional to $v_\Delta$. 
For example, $v_\Delta$ might be determined via the measurement of the $H^\pm W^\mp Z$ vertex at the LHC~\cite{HWZ_LHC} 
or its luminosity upgraded version and at the ILC~\cite{HWZ_ILC}. 
The other vertices such as $H^{\pm\pm}W^\mp W^\mp$~\cite{Kadastik:2007yd} and $H^{\pm\pm}H^\mp H^\mp$ etc. can also be used to measure $v_\Delta$: 
see Appendix~C. 

\section{Conclusions}
We have calculated radiative corrections to the electroweak observables in the $Y=1$ HTM. 
Such a model is motivated by the scenario of generating the neutrino masses via the type II seesaw mechanism. 
In this model, the rho parameter deviates from unity at the tree level, so that  
the electroweak sector is described by the four input parameters such as 
$\alpha_{\text{em}}$, $G_F$, $m_Z$ and $\sin^2\theta_W$.  
We have evaluated  one-loop contributions to the rho parameter and the W boson mass, 
and we have examined the possible mass spectrum of the extra Higgs bosons under the constraint from the electroweak precision data. 
We have found that the hierarchical mass spectrum among $H^{\pm\pm}$, $H^{\pm}$ and $A$ (or $H$) 
can be allowed without contradiction with the data 
especially in Case I ($m_A$ $(\simeq m_H)>m_{H^+}>m_{H^{++}}$). 

In our analysis, the mass of $H^{\pm\pm}$ is of $\mathcal{O}(100-200)$ GeV, with $\Delta m$ to be a few hundred GeV and $v_\Delta$ of 
several GeV are preferred by the electroweak precision data. 

Furthermore, regarding the recently Higgs boson searches at the LHC, 
we have discussed a few phenomenological consequences of such a mass spectrum with 
relatively large mass splitting under the constraint from the electroweak precision data.
\\

$Acknowledgments$

The authors would like to Mayumi Aoki, Mariko Kikuchi and Hiroyuki Taniguchi for useful discussions. 
The work of S.K. was supported in part by Grant-in-Aid for Scientific Research, Nos. 22244031 and 23104006. 
K.Y. was supported by Japan Society for the Promotion of Science.

\appendix
\section{Gauge boson self-energies}
In this Appendix, 
analytic expressions for the gauge boson self-energies at the one-loop level are listed 
in terms of the Passarino-Veltman functions~\cite{Passarino:1978jh}. 
The fermionic-loop contributions to the transverse part of the gauge boson two point functions are 
given in Ref.~\cite{hhkm}. 
The bosonic-loop contributions are listed below. The W boson two-point function is calculated as
\begin{align}
&\left(\frac{1}{16\pi^2}\right)^{-1}\Pi_{T}^{WW}(p^2)
=\left(\frac{1}{16\pi^2}\right)^{-1}\Pi_{T}^{WW}(p^2)_\text{SM}\notag\\
&+g^4\left(\frac{v_\Phi}{2}c_\alpha+v_\Delta s_\alpha\right)^2B_0(p^2,m_h,m_W)
+g^4\left(-\frac{v_\Phi}{2}s_\alpha+v_\Delta c_\alpha\right)^2B_0(p^2,m_H,m_W)\notag\\
&+2g^4v_\Delta^2B_0(p^2,m_{H^{++}},m_W)\notag\\
&+\frac{g^4}{2\hat{c}_W^2}v_\Delta^2c_{\beta_\pm}^2B_0(p^2,m_{H^+},m_W)
+\frac{g^4}{\hat{c}_W^2}\left[\frac{v_\Phi}{2}\hat{s}_W^2c_{\beta_\pm}
+\frac{v_\Delta}{\sqrt{2}}(1+\hat{s}_W^2)s_{\beta_\pm}\right]^2B_0(p^2,m_Z,m_W)\notag\\
&+\frac{g^2e^2}{4}(v_\Phi^2+2v_\Delta^2)B_0(p^2,0,m_W)\notag\\
&+g^2c_{\beta_\pm}^2B_5(p^2,m_{H^{++}},m_{H^+})+g^2s_{\beta_\pm}^2B_5(p^2,m_{H^{++}},m_W)\notag\\
&+\frac{g^2}{4}(c_\alpha s_{\beta_\pm}-\sqrt{2}s_\alpha c_{\beta_\pm})^2B_5(p^2,m_{H^+},m_h)
+\frac{g^2}{4}(c_\alpha c_{\beta_\pm}+\sqrt{2}s_\alpha s_{\beta_\pm})^2B_5(p^2,m_W,m_h)\notag\\
&+\frac{g^2}{4}(s_\alpha s_{\beta_\pm}+\sqrt{2}c_\alpha c_{\beta_\pm})^2B_5(p^2,m_{H^+},m_H)
+\frac{g^2}{4}(s_\alpha c_{\beta_\pm}-\sqrt{2}c_\alpha s_{\beta_\pm})^2B_5(p^2,m_W,m_H)\notag\\
&+\frac{g^2}{4}(s_{\beta_0} s_{\beta_\pm}+\sqrt{2}c_{\beta_0} c_{\beta_\pm})^2B_5(p^2,m_{H^+},m_A)
+\frac{g^2}{4}(s_{\beta_0} c_{\beta_\pm}-\sqrt{2}c_{\beta_0} s_{\beta_\pm})^2B_5(p^2,m_W,m_A)\notag\\
&+\frac{g^2}{4}(-c_{\beta_0} s_{\beta_\pm}+\sqrt{2}s_{\beta_0} c_{\beta_\pm})^2B_5(p^2,m_{H^+},m_Z)
+\frac{g^2}{4}(c_{\beta_0} c_{\beta_\pm}+\sqrt{2}s_{\beta_0} s_{\beta_\pm})^2B_5(p^2,m_W,m_Z), 
\end{align}
where the function $B_5$ is given as~\cite{hhkm}
\begin{align}
B_5(p^2,m_1,m_2)&=A(m_1)+A(m_2)-4B_{22}(p^2,m_1,m_2).
\end{align}
The photon two-point function is calculated as
\begin{align}
\left(\frac{1}{16\pi^2}\right)^{-1}\Pi_{T}^{\gamma\gamma}(p^2)&=
\left(\frac{1}{16\pi^2}\right)^{-1}\Pi_{T}^{\gamma\gamma}(p^2)_\text{SM}\notag\\
&+\frac{e^2g^2}{2}(v_\Phi^2+2v_\Delta^2)B_0(p^2,m_W,m_W)\notag\\
&+4e^2B_5(p^2,m_{H^{++}},m_{H^{++}})+e^2B_5(p^2,m_{H^+},m_{H^+})+e^2B_5(p^2,m_W,m_W). 
\end{align}
The photon-Z boson mixing is calculated as
\begin{align}
\left(\frac{1}{16\pi^2}\right)^{-1}\Pi_{T}^{\gamma Z}(p^2)
&=\left(\frac{1}{16\pi^2}\right)^{-1}\Pi_{T}^{\gamma Z}(p^2)_\text{SM}\notag\\
&+g^4\frac{\hat{s}_W}{\hat{c}_W}\sqrt{v_\Phi^2+2v_\Delta^2}\left[\frac{v_\Phi}{2}\hat{s}_W^2c_{\beta_\pm}+\frac{v_\Delta}{\sqrt{2}}(1+\hat{s}_W^2)s_{\beta_\pm}\right]
B_0(p^2,m_W,m_W)\notag\\
&-2g^2\frac{\hat{s}_W(\hat{c}_W^2-\hat{s}_W^2)}{\hat{c}_W}B_5(p^2,m_{H^{++}},m_{H^{++}})
-\frac{g^2}{2}\frac{\hat{s}_W}{\hat{c}_W}(\hat{c}_W^2-\hat{s}_W^2-c_{\beta_\pm}^2)B_5(p^2,m_{H^+},m_{H^+})\notag\\
&-\frac{g^2}{2}\frac{\hat{s}_W}{\hat{c}_W}(\hat{c}_W^2-\hat{s}_W^2-s_{\beta_\pm}^2)B_5(p^2,m_W,m_W),
\end{align}
where $\Pi_{T}^{VV}(p^2)_\text{SM}$ are the SM gauge boson loop contributions. 
These are calculated as 
\begin{align}
\left(\frac{1}{16\pi^2}\right)^{-1}\Pi_{T}^{WW}(p^2)_\text{SM}&=
-g^2\hat{s}_W^2[6(D-1)B_{22}+p^2(2B_{21}+2B_1+5B_0)](p^2,0,m_W)\notag\\
&-g^2\hat{c}_W^2[6(D-1)B_{22}+p^2(2B_{21}+2B_1+5B_0)](p^2,m_Z,m_W)\notag\\
&+g^2(D-1)\left[\hat{c}_W^2A(m_Z)+A(m_W)\right]\notag\\
&+2e^2\left[B_{22}(p^2,0,m_W)+\frac{\hat{c}_W^2}{\hat{s}_W^2}B_{22}(p^2,m_Z,m_W)\right]\notag\\
&-4g^2(p^2-m_W^2)[\hat{c}_W^2B_0(p^2,m_W,m_Z)+\hat{s}_W^2B_0(p^2,m_W,0)]
,\\
\left(\frac{1}{16\pi^2}\right)^{-1}\Pi_{T}^{\gamma \gamma}(p^2)_\text{SM}&=
-e^2[6(D-1)B_{22}+p^2(2B_{21}+2B_1+5B_0)](p^2,m_W,m_W)\notag\\
&+2e^2(D-1)A(m_W)+2e^2B_{22}(p^2,m_W,m_W)\notag\\
&-4e^2p^2B_0(p^2,m_W,m_W),\\
\left(\frac{1}{16\pi^2}\right)^{-1}\Pi_{T}^{\gamma Z}(p^2)_\text{SM}&=
+e^2\frac{\hat{c}_W}{\hat{s}_W}[6(D-1)B_{22}+p^2(2B_{21}+2B_1+5B_0)](p^2,m_W,m_W)\notag\\
&-2e^2\frac{\hat{c}_W}{\hat{s}_W}(D-1)A(m_W),\notag\\
&-2e^2\frac{\hat{c}_W}{\hat{s}_W}B_{22}(p^2,m_W,m_W)\notag\\
&+4g^2\frac{\hat{s}_W}{\hat{c}_W}\left(p^2-\frac{m_W^2}{2}\right)B_0(p^2,m_W,m_W),
\end{align}
where $D=4-2\epsilon$.

\section{Feynman Rules}
We here list the Feynman rules for gauge-scalar interactions in the Table~\ref{fr1}. 

\begin{table}[!h]
\begin{center}
{\renewcommand\arraystretch{1.5}
\begin{tabular}{|c||c|}\hline
Vertices& Gauge couplings \\\hline
$H^{++}W_\mu^-W_\nu^- $ & $ig^2\frac{v_\Delta}{\sqrt{2}}g_{\mu\nu}$\\\hline
$H^{++}H^-W_\mu^- $ & $ig\cos\beta_\pm(p_1-p_2)_\mu$\\\hline
$H^{++}w^-W_\mu^- $ & $ig\sin\beta_\pm(p_1-p_2)_\mu$\\\hline
$H^+W_\mu^-Z_\nu $ & $-i\frac{g^2}{\hat{c}_W}\frac{v_\Delta}{\sqrt{2}}\cos\beta_\pm g_{\mu\nu}$\\\hline
$w^+W_\mu^-Z_\nu $ & $-i\frac{g^2}{\hat{c}_W}\left[ \frac{v}{2}\hat{s}_W^2\cos\beta_\pm+\frac{v_\Delta}{\sqrt{2}}(1+\hat{s}_W^2)\sin\beta_\pm  \right]g_{\mu\nu}$\\\hline
$H^+hW_\mu^- $ & $-i\frac{g}{2}(\cos\alpha\sin\beta_\pm-\sqrt{2}\sin\alpha\cos\beta_\pm)(p_1-p_2)_\mu$\\\hline
$H^+HW_\mu^- $ & $i\frac{g}{2}(\sin\alpha\sin\beta_\pm+\sqrt{2}\cos\alpha\cos\beta_\pm)(p_1-p_2)_\mu$\\\hline
$H^+AW_\mu^- $ & $\frac{g}{2}(\sin\beta_0\sin\beta_\pm+\sqrt{2}\cos\beta_0\cos\beta_\pm)(p_1-p_2)_\mu$\\\hline
$H^+zW_\mu^- $ & $\frac{g}{2}(-\cos\beta_0\sin\beta_\pm+\sqrt{2}\sin\beta_0\cos\beta_\pm)(p_1-p_2)_\mu$\\\hline
$w^+hW_\mu^- $ & $-i\frac{g}{2}(-\cos\alpha\cos\beta_\pm-\sqrt{2}\sin\alpha\sin\beta_\pm)(p_1-p_2)_\mu$\\\hline
$w^+HW_\mu^- $ & $i\frac{g}{2}(\sin\alpha\cos\beta_\pm-\sqrt{2}\cos\alpha\sin\beta_\pm)(p_1-p_2)_\mu$\\\hline
$w^+AW_\mu^- $ & $-\frac{g}{2}(\sin\beta_0\cos\beta_\pm-\sqrt{2}\cos\beta_0\sin\beta_\pm)(p_1-p_2)_\mu$\\\hline
$w^+zW_\mu^- $ & $\frac{g}{2}(\cos\beta_0\cos\beta_\pm+\sqrt{2}\sin\beta_0\sin\beta_\pm)(p_1-p_2)_\mu$\\\hline
$AhZ_\mu $ & $\frac{g}{2\hat{c}_W}(-\cos\alpha\sin\beta_0+2\sin\alpha\cos\beta_0)(p_1-p_2)_\mu$\\\hline
$H^{++}H^{--}A_\mu $ & $2ie(p_1-p_2)_\mu$\\\hline
$H^{+}H^{-}A_\mu $ & $ie(p_1-p_2)_\mu$\\\hline
$w^{+}w^{-}A_\mu $ & $ie(p_1-p_2)_\mu$\\\hline
$H^{++}H^{--}Z_\mu $ & $ig\frac{\hat{c}_W^2-\hat{s}_W^2}{\hat{c}_W}(p_1-p_2)_\mu$\\\hline
$H^{+}H^{-}Z_\mu $ & $i\frac{g}{2\hat{c}_W}(\hat{c}_W^2-\hat{s}_W^2-c_{\beta_\pm}^2)(p_1-p_2)_\mu$\\\hline
$w^{+}w^{-}Z_\mu $ & $i\frac{g}{2\hat{c}_W}(\hat{c}_W^2-\hat{s}_W^2-s_{\beta_\pm}^2)(p_1-p_2)_\mu$\\\hline
\end{tabular}
\hspace{3mm}
\begin{tabular}{|c||c|}\hline
Vertices& Gauge couplings \\\hline
$H^{++}H^{--}A_\mu A_\nu $ & $4ie^2g_{\mu\nu}$\\\hline
$H^{+}H^{-}A_\mu A_\nu $ & $ie^2g_{\mu\nu}$\\\hline
$w^{+}w^{-}A_\mu A_\nu $ & $ie^2g_{\mu\nu}$\\\hline
$H^{++}H^{--}W_\mu^+ W_\nu^- $ & $ig^2g_{\mu\nu}$\\\hline
$H^{+}H^{-}W_\mu^+ W_\nu^- $ & $i\frac{g^2}{4}(5+3\cos2\beta_\pm)g_{\mu\nu}$\\\hline
$w^{+}w^{-}W_\mu^+ W_\nu^- $ & $i\frac{g^2}{4}(5-3\cos2\beta_\pm)g_{\mu\nu}$\\\hline
$HHW_\mu^+ W_\nu^- $ & $i\frac{g^2}{8}(3+\cos2\alpha)g_{\mu\nu}$\\\hline
$hhW_\mu^+ W_\nu^- $ & $i\frac{g^2}{8}(3-\cos2\alpha)g_{\mu\nu}$\\\hline
$AAW_\mu^+ W_\nu^- $ & $i\frac{g^2}{8}(3+\cos2\beta_0)g_{\mu\nu}$\\\hline
$zzW_\mu^+ W_\nu^- $ & $i\frac{g^2}{8}(3-\cos2\beta_0)g_{\mu\nu}$\\\hline
$H^{++}H^{--}A_\mu Z_\nu $ & $i4g^2\frac{\hat{s}_W}{\hat{c}_W}(\hat{c}_W^2-\hat{s}_W^2)g_{\mu\nu}$\\\hline
$H^+H^-A_\mu Z_\nu $ & $ig^2\frac{\hat{s}_W}{\hat{c}_W}(\hat{c}_W^2-\hat{s}_W^2-\cos^2\beta_\pm)g_{\mu\nu}$\\\hline
$w^+w^-A_\mu Z_\nu $ & $ig^2\frac{\hat{s}_W}{\hat{c}_W}(\hat{c}_W^2-\hat{s}_W^2-\sin^2\beta_\pm)g_{\mu\nu}$\\\hline
\end{tabular}}
\caption{Feynman rules for scalar bosons and gauge bosons three-pont and four-point interaction.}
\label{fr1}
\end{center}
\end{table}


\end{document}